\documentclass[twocolumn]{aastex631}
\usepackage{amsmath}
\usepackage{wasysym}
\usepackage{graphicx}
\usepackage{amssymb}
\usepackage{epstopdf}
\usepackage{mathrsfs}
\usepackage{anyfontsize}
\usepackage{natbib}
\usepackage{CJK}
\usepackage{url}
\usepackage{color}
\usepackage{lipsum}
\usepackage{diagbox}
\usepackage{gensymb}
\usepackage{appendix}
\usepackage{booktabs}
\usepackage{threeparttable}
\usepackage[figuresright]{rotating}
\usepackage{multirow}

\shorttitle{}
\shortauthors{Hu et al.}
\begin{document}

\begin{CJK*}{UTF8}{gbsn}

	\title{Identification of BHB stars using Synthetic SkyMapper colors from Gaia XP spectra}

\author[0000-0003-1828-5318]{Guozhen Hu}
\altaffiliation{Physics Postdoctoral Research Station at Hebei Normal University}
 \affiliation{Department of Physics, Hebei Normal University, Shijiazhuang 050024, People's Republic of China}
 \affiliation{Guo Shoujing Institute for Astronomy, Hebei Normal University, Shijiazhuang 050024, People's Republic of China}
 \affiliation{Hebei Key Laboratory of Photophysics Research and Application, Shijiazhuang 050024, People's Republic of China}
 
 \author[0000-0001-8424-1079]{Yang Huang}
\affiliation{School of Astronomy and Space Science, University of Chinese Academy of Sciences, Beijing 100049, People's Republic of China}
\affiliation{National Astronomical Observatories, Chinese Academy of Sciences, Beijing 100012, People's Republic of China}
\affiliation{Institute for Frontiers in Astronomy and Astrophysics, Beijing Normal University, Beijing, 102206, People's Republic of China}

\author[0000-0003-1359-9908]{Wenyuan Cui}
\affiliation{Department of Physics, Hebei Normal University, Shijiazhuang 050024, People's Republic of China}
\affiliation{Guo Shoujing Institute for Astronomy, Hebei Normal University, Shijiazhuang 050024, People's Republic of China}

\author{Tao Wang}
\affiliation{South-Western Institute for Astronomy Research, Yunnan University, Kunming, 650500, P. R. China}

\author[0000-0001-8424-1079]{Kai Xiao}
\affiliation{School of Astronomy and Space Science, University of Chinese Academy of Sciences, Beijing 100049, People's Republic of China}

\author{Ruifeng Shi}
\affiliation{School of Astronomy and Space Science, University of Chinese Academy of Sciences, Beijing 100049, People's Republic of China}

\author{Jie Ju}
\affiliation{Department of Physics, Hebei Normal University, Shijiazhuang 050024, People's Republic of China}
\affiliation{School of Sciences, Hebei University of Science and Technology, Shijiazhuang 050018, People's Republic of China}

\author[0000-0002-1259-0517]{Bowen Huang}
\affiliation{Institute for Frontiers in Astronomy and Astrophysics, Beijing Normal University, Beijing, 102206, China}
\affiliation{School of Physics and Astronomy, Beijing Normal University No.19, Xinjiekouwai St, Haidian District, Beijing, 100875, China}
\author[0000-0003-2021-4818]{Chunyan Li}
\affiliation{School of Physics and Astronomy, China West Normal University, No. 1 Shida Road, Nanchong 637002, People's Republic of China
}

\author[0000-0001-5314-2924]{Zhicun Liu}
\altaffiliation{Physics Postdoctoral Research Station at Hebei Normal University}
\affiliation{Department of Physics, Hebei Normal University, Shijiazhuang 050024, People's Republic of China}
\affiliation{Guo Shoujing Institute for Astronomy, Hebei Normal University, Shijiazhuang 050024, People's Republic of China}
\affiliation{Hebei Key Laboratory of Photophysics Research and Application, Shijiazhuang 050024, People's Republic of China}

\author[0000-0003-2536-3142]{Xiaolong Wang}
\altaffiliation{Physics Postdoctoral Research Station at Hebei Normal University}
\affiliation{Department of Physics, Hebei Normal University, Shijiazhuang 050024, People's Republic of China}
\affiliation{Guo Shoujing Institute for Astronomy, Hebei Normal University, Shijiazhuang 050024, People's Republic of China} 
\affiliation{Hebei Advanced Thin Films Laboratory, Shijiazhuang 050024, People's Republic of China}
\author{Changqing Luo}
\affiliation{National Astronomical Observatories, Chinese Academy of Sciences, Beijing 100012, People's Republic of China}

\correspondingauthor{Yang Huang and Wenyuan Cui}
\email{huangyang@ucas.ac.cn, cuiwenyuan@hebtu.edu.cn}

\begin{abstract}
Blue horizontal-branch (BHB) stars are ideal tracers for mapping the structure of Galactic stellar halo. Traditionally, BHB sample stars are built from large-scale spectroscopic surveys utilizing their spectral features, however, the resulting sample sizes have been quite limited. In this paper, we construct a catalog of BHB stars based on synthetic colors $(u-v)_{0}$ and $(g-i)_{0}$ in SkyMapper photometric systems, which are convolved from Gaia XP spectra. A total of 49,733 BHB stars are selected from nearly the entire sky (excluding regions of low Galactic latitudes $|b| \le 8^{\circ}$ with heavy reddening), with a completeness and purity exceeding 90\%. Using member stars of globular clusters with precise distance determinations, we carefully calibrate the relationship between the $g$-band absolute magnitude and  $(g-i)_{0}$, achieving a precision of 0.11\,mag, which corresponds to a 5\% uncertainty in distance. This relation is applied to derive distances for all BHB stars in the constructed sample. Given current capabilities of Gaia XP observations, the constructed BHB sample is primarily located within 20 kpc, enabling detailed mapping of the inner stellar halo. To extend this depth to the outer halo or even the edge of our Galaxy, we explore the potential of the Chinese Space Station Telescope (CSST) and its broad-band photometry for detecting BHB stars. Using mock data from synthetic spectra, we find that it is feasible to distinguish BHB stars from blue stragglers (BS) stars using CSST near-ultraviolet bands ($NUV, u$) photometry. Thanks to the deep limiting magnitude of CSST, its data will provide a groundbreaking perspective on our Galaxy, particularly regarding the outer halo, in an unprecedented volume.

\end{abstract}
	

\section{Introduction}\label{sec:intro}

Blue horizontal-branch (BHB) stars are horizontal-branch stars located on the blue side of RR Lyrae instability strip on the Hertzsprung-Russell diagram (see the review of \citealt{2009Ap&SS.320..261C}). They are helium-burning giants with mass range of approximately 0.5 - 1.0 $M_{\odot}$ \citep{2010ApJ...708L.121D}. As old and metal-poor tracers, BHB stars are often found in globular clusters and the Galactic stellar halo \citep{1952AJ.....57....4A,1955ApJS....2....1H,2001PASP..113.1162M,2015ApJ...813L..16S}. \\

BHB stars are well-known ``standard candle" due to their nearly constant absolute magnitudes within a restricted color range. Their relatively high luminosity enables observations that encompass a large volume of the Galaxy \citep{1974ApJS...28..157G,2004AJ....127..899S}.
In this regard, BHB stars are ideal tracers for studying kinematics and structural composition of the Milky Way (MW; e.g. \citealt{1974ApJS...28..157G,1984ApJ...281..260P,2000ApJ...540..825Y}). 
Due to their greater abundance compared to RR Lyrae stars, BHB stars are preferred for investigating various aspects of the Galactic halo (\citealt{1998MNRAS.297..732S,2008ApJ...684.1143X,2010ApJ...708L.121D,2011MNRAS.416.2903D,2015ApJ...813L..16S}). 
For example, using a catalog of BHB stars from the Sloan Digital Sky Survey/Sloan Extension for
Galactic Understanding and Exploration (SDSS/SEGUE, \citealt{2000AJ....120.1579Y}), \cite{2008ApJ...684.1143X} estimated the mass of our Galaxy. \cite{2020MNRAS.499.1058U} determined the distance to the Galactic center utilizing a catalog of 4985 BHB stars compiled by \cite{2011ApJ...738...79X}. \\

In general, samples of BHB stars are constructed from large-scale spectroscopic surveys that utilize their spectral features, including the strong Balmer jump and the narrow, prominent Balmer-line profiles (e.g. \citealt{2002MNRAS.337...87C,2008ApJ...684.1143X,2012AJ....143...86V}). Through color cuts from SDSS photometry and selection criteria based on Balmer-line profile, \cite{2011ApJ...738...79X} obtained 4985 BHB stars from SDSS DR8. Most recently, \cite{2024ApJS..270...11J} identified about 5000 BHB stars by combining the spectral indices and a set of Balmer line profile selection criteria from the Large Sky Area Multi-Object Fiber Spectroscopic Telescope (LAMOST) DR5. However, the number of spectroscopically selected sample stars is quite limited (no more than ten thousand), and these samples are also affected by significant selection biases. \\

Alternatively, photometry-based methods have been developed for identifying BHB stars without the need for spectral observations, offering greater efficiency. The core idea of this method lies in the sensitivity of the photometric color,  typically measured using blue filters, to the Balmer jump.
In practice, this method typically identifies BHB stars by selecting them from the blue stragglers (BS) stars based on the stellar color locus, with a focus on choosing the appropriate colors. 
This approach has been widely applied: \cite{1991ApJ...375..121P} selected BHB stars from other A-type stars in $(U - B)_{0}$ versus $(B - V)_{0}$ space. \cite{2000ApJ...540..825Y} identified 1493 BHB stars using a color cut in the $u-g$ versus $g-r$ color-color diagram. Using deep $u$-band imaging from the new Canada-France Imaging Survey (CFIS) in combination with $griz$ bands from Pan-STARRS 1, \cite{2018MNRAS.481.5223T} constructed a BHB catalog. However, these studies utilized broad photometric bands, which are not effective in distinguishing BHB stars from BS stars. When relying solely on broad-band information, the success rates for differentiating between BHB and BS stars have demonstrated completeness levels ranging from 50$\% $ to 70$\%$, while contamination rates may be as high as 30$\%$ (e.g., \citealt{2010AJ....140.1850B,2018PASJ...70...69F,2018MNRAS.481.5223T}), which hinders efforts to accurately map the structure of the stellar halo.\\

Recently, using SkyMapper Southern Sky Survey (SMSS) DR4 photometry with medium $u$- and $v$- filters, T. Wang et al. (to be submitted) constructed a high-quality BHB sample with purity and completeness both higher than 90 per cent, containing more than 80,000 BHB stars. However, those BHB stars were limited to the southern sky (SMSS fields). Fortunately, Gaia DR3 data includes BP/RP (XP) spectra from approximately 200 million sources distributed across the sky, with G band brighter than 17.65 mag, which nearly covers the wavelength ranges of the $u$ and $v$ filters used in SMSS. This dataset enables the identification of BHB stars across the entire sky based on synthetic stellar colors derived from Gaia XP spectra. More recently, \cite{2024ApJS..271...13H} accurately measured and corrected systematic errors related to magnitude, color, and extinction in the Gaia DR3 XP spectra (hereafter ``corrected" Gaia XP spectra).
The systematic errors in the blue range of the Gaia DR3 XP spectra have been significantly reduced, providing a solid foundation for this work.\\

In this paper, we aim to construct a high-quality, all-sky distributed bright BHB sample based on the synthetic SMSS $uv$ photometry utilizing the ``corrected" Gaia XP spectra.
In the future, we aim to extend our research beyond the inner halo and investigate the outer edge of our Galaxy using the BHB sample. To this end, we will explore the capabilities of the Chinese Space Station Telescope (CSST) for detecting BHB stars, leveraging its extensive sky coverage and survey depth.
This paper is structured as follows.
In Section 2, we describe the data used in this work and our selection criteria for BHB stars. In Section 3, we determine the distances of our BHB stars by calibrating their absolute magnitudes. In Section 4, we discuss the spatial distribution of our BHB stars and explore the ability of CSST near-ultraviolet bands ($NUV, u$) photometry to select BHB stars. Finally, we summarize our results in Section 5.

\section{Data and Identification of BHB stars}\label{sec:data}

\subsection{Synthetic SkyMapper Photmetry}

The synthetic magnitudes and their errors for approximately 200 million stars of the six-bands in the SMSS system were initially compiled and were constructed using an improved synthetic photometry method \citep{2023ApJS..269...58X} based on the ``corrected" Gaia DR3 XP spectra \citep{2024ApJS..271...13H}. 
This SMSS standard catalog is included in the BEst STar (BEST) database (Xiao et al. to be submitted) and will be publicly available. \\

This work considers only stars satisfying the following criteria:

\begin{enumerate}
\item[(a)] \texttt{FLAG} = 000/001/002 to ensure high-quality ``corrected" Gaia XP spectra \citep{2024ApJS..271...13H}.
\item[(b)] Magnitude errors in the $uvgi$ bands are required to be less than 0.04 to maintain high photometric precision. 
\item[(c)] $\lvert b \rvert \geq$ 8$^{\circ}$ to avoid regions of high reddening \footnote{Typical value of $E(B-V)$ at $\lvert b \rvert $ = 8$^{\circ}$ is aroud 0.33 mag. In such a relatively low extinction (thus low extinction correction errors), BHB and BS can still be well separated.} and ensure reliable $E(B-V)$ estimates from the reddening map of \cite{1998ApJ...500..525S}. 
\end{enumerate}

Finally, a total of 29,232,880 stars are left (hereafter parent sample).

\begin{figure} 
	\centering
	\includegraphics[width=0.5\textwidth]{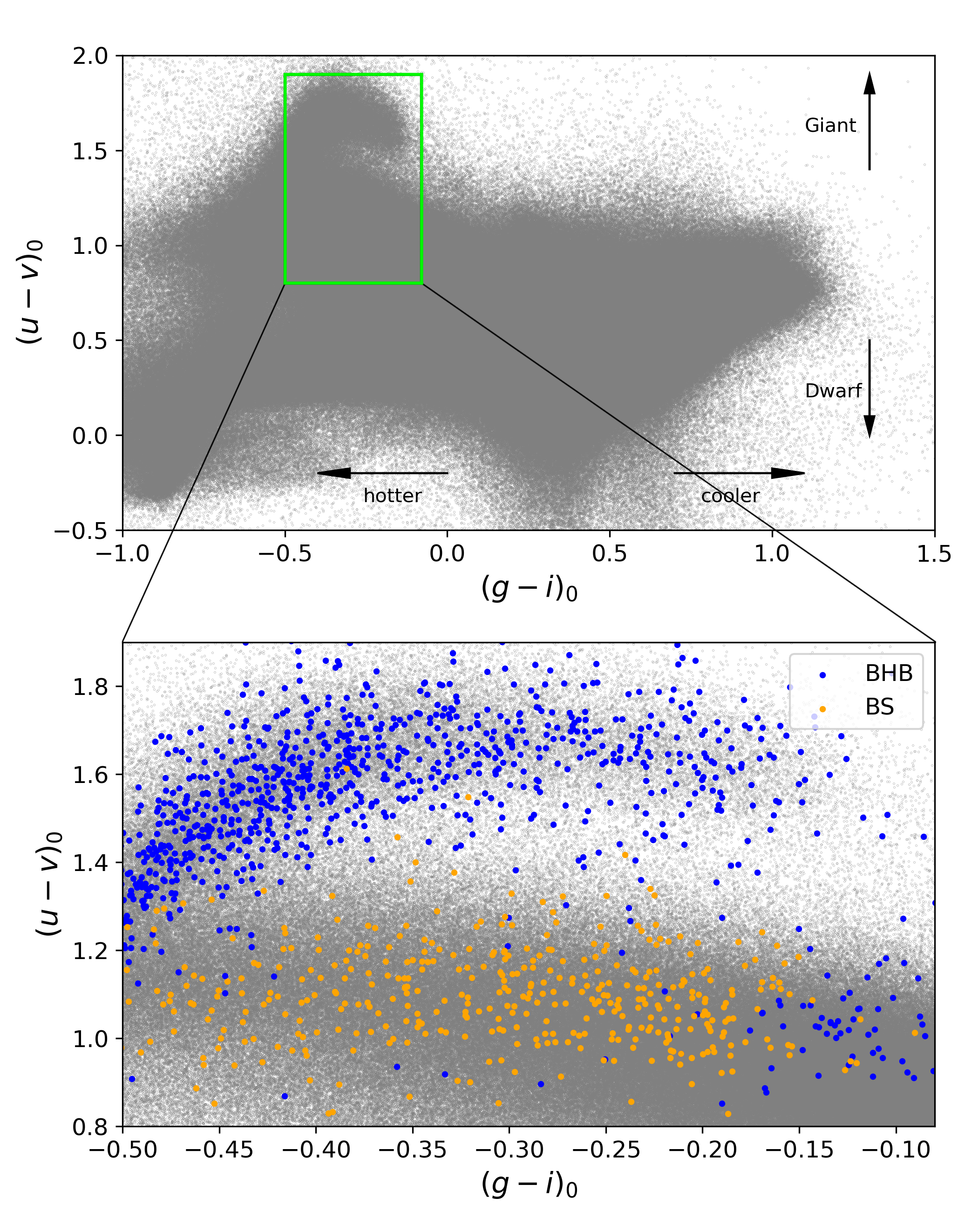} 
	\caption{Top panel: the distribution of stars of parent sample  in $(u-v)_{0}$ vs $(g-i)_{0}$ plane. Bottom panel: this panel magnifies the green box ($0.8 \le (u-v)_{0} \le 1.9$ and $-0.50 \le (g-i)_{0} \le -0.08$) of top panel to show the distinct difference in the distribution of BHB and BS. Orange and blue dots represent BS stars and BHB stars, which come from \cite{2008ApJ...684.1143X} and \cite{2011ApJ...738...79X}, respectively.}
	\label{Fig:1}
\end{figure}

\subsection{ Identification of BHB Stars from color-color diagram}\label{sec:2.2}

Following the technique developed by T. Wang et al. (to be submitted), we select BHB stars in the $(u-v)_{0}$ versus $(g-i)_{0}$ space.
The $u$ band shows sensitivities to both surface gravity log $\text{\sl g}$ and metallicity. Meanwhile, $v$ band is mainly influenced by metallicity. In this regard, the combination of $(u-v)_{0}$ serves as a reliable proxy for log $\text{\sl g}$. The redder $(u-v)_{0}$ color is indicative of a lower log $\text{\sl g}$, whereas bluer $(u-v)_{0}$ corresponds to higher log $\text{\sl g}$. 
On the other hand, the color $(g-i)_{0}$ is a well-established indicator of effective temperature. Considering the distinct log $\text{\sl g}$ values for BHB and BS stars (BHB: log $\text{\sl g}$ $\sim$ 2.8-3.75, BS: log $\text{\sl g}$ $\sim$ 3.75-5.0; \citealt{2002MNRAS.337...87C,2008AJ....135..564B}), we expect a clear distinction in their distribution on the $(u-v)_{0}$ versus $(g-i)_{0}$ color-color plane. We therefore show all the stars of parent sample in the $(u-v)_{0}$ versus $(g-i)_{0}$ (see the) top panel of Figure\,\ref{Fig:1}.
The colors $(u-v)_{0}$ and $(g-i)_{0}$ have been corrected for reddening using $E(B-V)$ values from the 2D dust reddening map by \cite{1998ApJ...500..525S}, along with the reddening coefficients adopted from Table 2 of \cite{2021ApJ...907...68H}.\\

In the bottom panel of Figure~\ref{Fig:1}, we zoom in on the region containing BHB and BS stars, defined by $0.8 \le (u-v)_{0} \le 1.9$ and $-0.50 \le (g-i)_{0} \le -0.08$. Two distinct branches are clearly visible in the color-color diagram. Under the guidance of BHB and BS stars identified from SDSS spectra \citep{2008ApJ...684.1143X,2011ApJ...738...79X}, the upper branch is predominantly composed of BHB stars, while the lower branch is mainly BS stars, which are in excellent agreement with our expectations. However, in the lower branch where $(g - i)_0$ is redder than $-0.20$, a region typically associated with BS stars, a significant number of BHB stars are also present. This is because, in this region, the spectral index fails to effectively differentiate between BHB and BS stars based on their spectral features \citep[see Figure 5 of][]{2008ApJ...684.1143X}, making their classification challenging.\\

\begin{figure}[htbp] 
	\centering
	\includegraphics[width=0.5\textwidth]{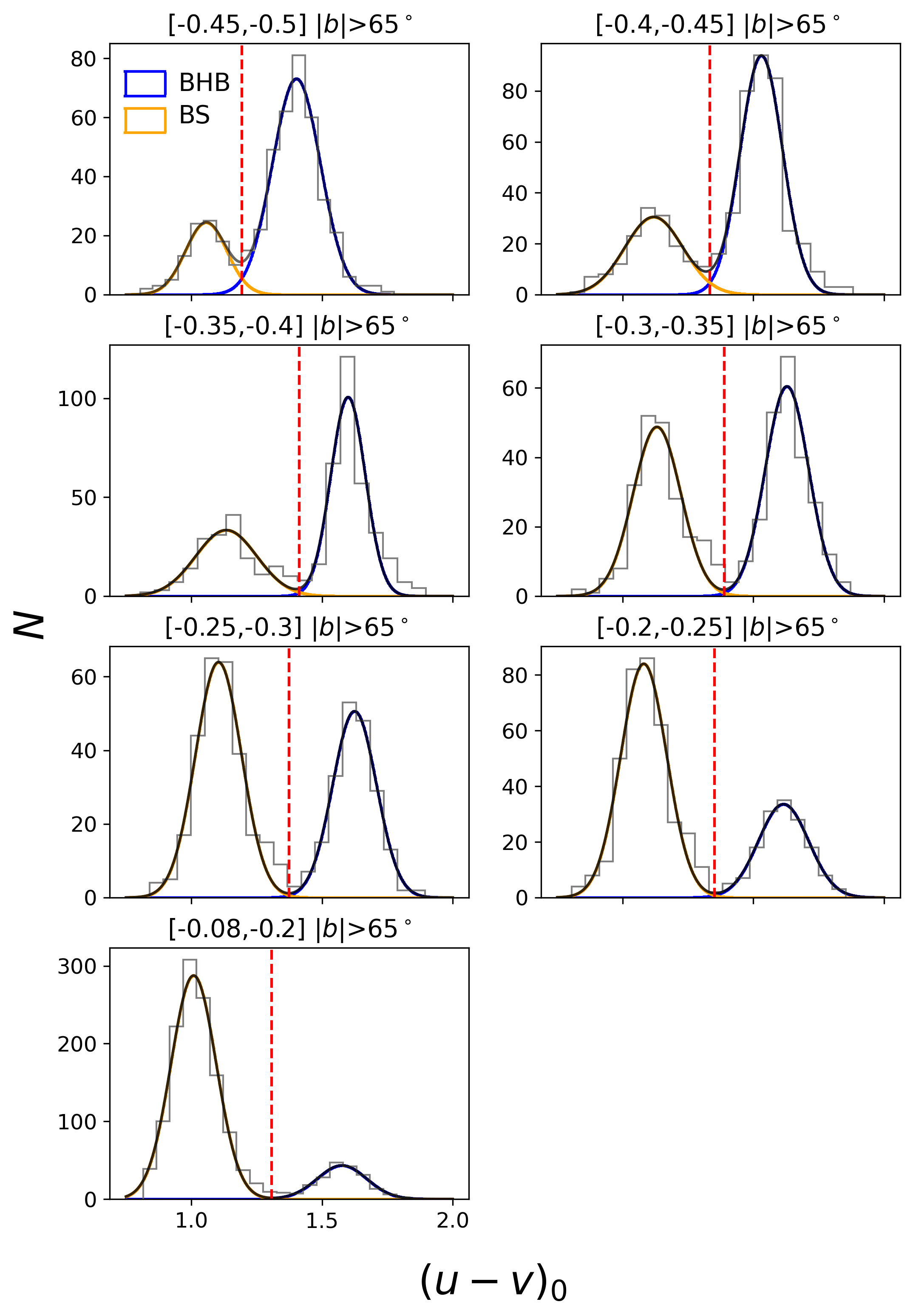} 
	\caption{The distribution of $(u-v)_{0}$ of stars with $\lvert b \rvert $ $>$ 65$^\circ$. The blue and orange lines represent the Gaussian distribution of BHB and BS stars, respectively. The black lines are the sum of two Gaussian distributions. The red dashed lines represent the intersection of two Gaussians distributions, delineating the boundary between the BHB and BS stars. 
    }
	\label{Fig:3}
\end{figure}

 To distinguish BHB stars from BS stars, we divide the stars in the box with $0.8 \le (u-v)_{0} \le 1.9 $ and $-0.50\le (g-i)_{0}\le -0.08$ into different bins based on their Galactic latitudes ($\lvert b \rvert $) and colors $(g-i)_{0}$.
Firstly, we divide the sample stars into seven bins based on Galactic latitude to ensure similar reddening values and a similar population across each bin . As expected, the difficulty in distinguishing between BHB and BS stars increases with higher reddening, which corresponds to lower Galactic latitudes.
For each Galactic latitude bin, stars are categorized into different groups based on their color  $(g-i)_{0}$. The detailed binning scheme can be found in Table~\ref{tab:B1}. As an example, we present the number distribution of stars in the $(u-v)_{0}$ space for $\lvert b \rvert $ $>$ 65$^\circ$ in Figure~\ref{Fig:3}.
The plot clearly shown two distinct peaks across all bins in $(g-i)_{0}$.
In this regard, we employ a two-Gaussian model to fit the number distribution along  $(u-v)_{0}$.
The fitting results are shown in Figure~\ref{Fig:3} for Galactic latitude bin $|b| > 65^{\circ}$ and the remaining Galactic latitude bins are shown in Figures~\ref{fig:side:a1}-\ref{fig:side:b3}, with the blue and orange lines represent the BHB and BS stars, respectively.
Here, we define the intersection of two Gaussian distributions as the boundary between the BHB and BS stars (see red dashed lines in each subplot of Figure~\ref{Fig:3} and Figures~\ref{fig:side:a1}-\ref{fig:side:b3}). 
Redder than the intersecting position but bluer than 1.9 are selected as final BHB candidates.
The number of candidate BHBs of each bin on Galactic latitude and stellar color $(g-i)_0$ is given in Table~\ref{tab:B1}.\\

\setcounter{table}{0} 
\setlength{\tabcolsep}{0pt}
\begin{sidewaystable}[h] 
\setlength{\abovecaptionskip}{0.05cm} 
\centering
\tiny

\caption{List of all newly identified BHB stars.} \label{tab:BHB}
\begin{threeparttable}

\begin{tabular*}{\hsize}{@{\extracolsep{\fill}}lrrrrrrrrrrrrrrrrrrl} 
\hline
\hline

 SOURCE-ID~\tnote{a} & ra & dec & pmra  & e\_pmra  & pmdec & e\_pmdec &u\_xpsp~\tnote{b} & u\_err~\tnote{b} &v\_xpsp ~\tnote{b}&  v\_err~\tnote{b} &g\_xpsp~\tnote{b} &  g\_err~\tnote{b} &i\_xpsp~\tnote{b} & i\_err~\tnote{b} & $E(B - V)$ ~\tnote{c}& dis (pc) &dis\_err (pc) &Cluster ~\tnote{d}\\
\hline

741357981454593536&154.06&29.31&$-$70.76&0.11&$-$22.84&0.14&7.08&0.19&5.79&0.13&5.41&0.02&5.84&0.02&0.03&84.04&0.073&$-$$-$\\
...& ... & ... & ... & ... & ... & ... & ... & ... & ... & ... & ...& ...& ...& ...& ...& ...& ...& ...\\
  4365635279142575488 & 254.29 & $-$4.07 & $-$4.71 &0.02& $-$6.45&0.02&16.51&0.23&14.72&0.04&14.05&0.01&13.88&0.01&0.29&3435.41&0.072&NGC\ 6254\\

\hline
\hline
\end{tabular*}
\tablecomments{The complete table is available on the publisher website.}
\begin{tablenotes}
		\item[a] Gaia DR3 source id.
 \item[b] From K. Xiao et al. (to be submitted).
 \item[c] Accepted from \cite{1998ApJ...500..525S}.
 \item[d] The name of the corresponding cluster.
\end{tablenotes}
\end{threeparttable}
\end{sidewaystable}

\subsection{Completeness and Purity of BHB stars}

To calculate the completeness and purity of our BHB sample, we define completeness as $C$, defined as
% \begin{center}
% $C = \frac{TP}{(TP + FN)}$~,\\
% \end{center}
\begin{equation}
C = \frac{\rm TP}{\rm TP + FN}~,
\end{equation}
where TP is ``true positive", i.e. BHB stars that are correctly identified as BHB stars. During the detailed calculation, TP represents the portion of the blue Gaussian's area that lies red-ward of the red dashed intersection line.

FN is ``false negative", i.e. the stars of BHB that are incorrectly identified as BS stars. FN represents the part area of the blue Gaussian that lies on the blue-ward side of the red-dash line.\\

As for the ``Purity" $P$, defined as:\\

\begin{equation}
P = \frac{\rm TP}{\rm TP + FP}~,\\
\end{equation} 

where TP is defined above, FP is ``false positive", which are BS stars misclassified as BHB stars. Here, FP refers to the portion of the orange Gaussian's area that lies red-ward of the red dashed intersection line.\\

\begin{figure}[htbp] 
	\centering
	\includegraphics[width=0.5\textwidth]{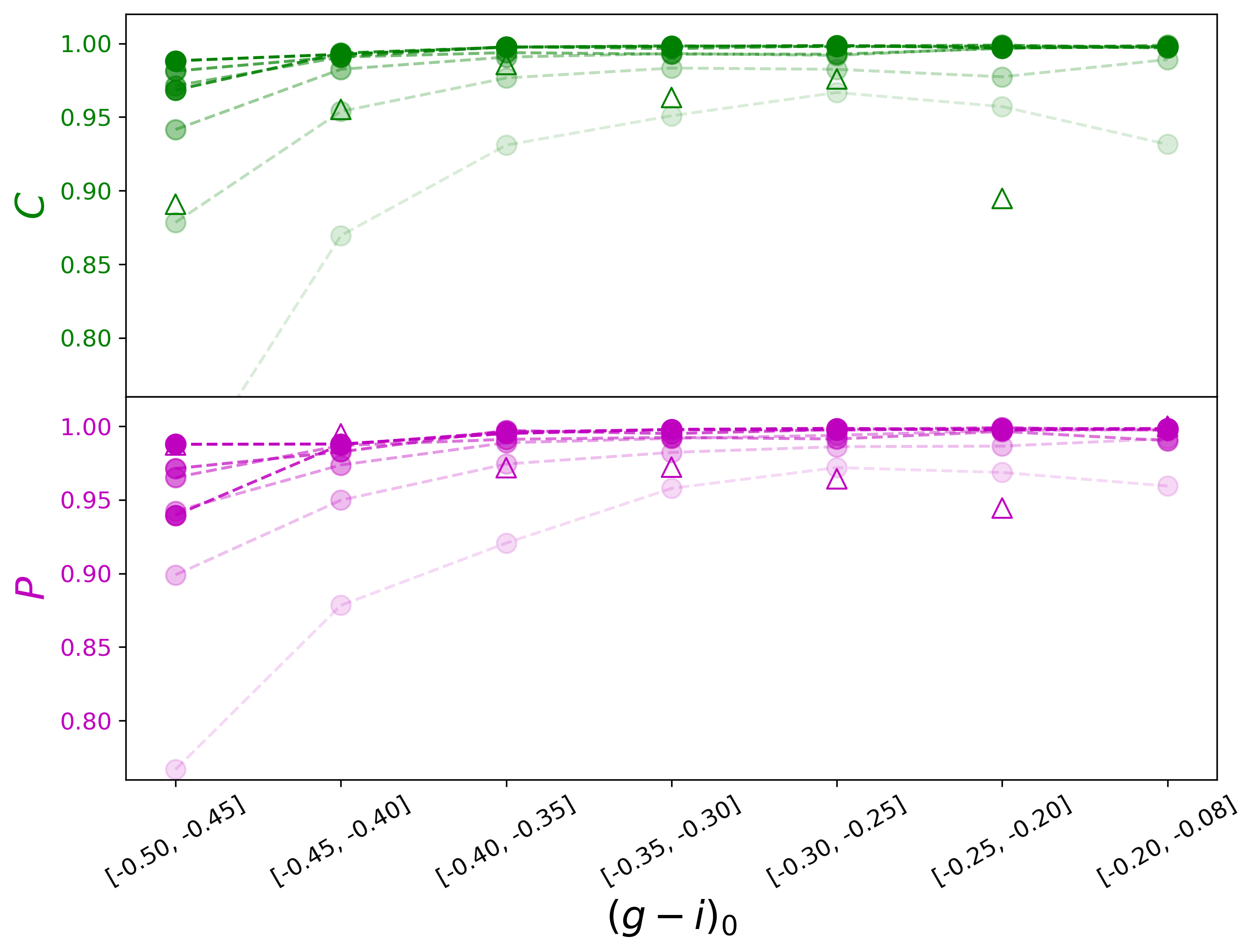} 
	\caption{The completeness (top panel) and purity (bottom panel) of identify BHB as a function of stellar color $(g-i)_{0}$. The hollow symbols represent \cite{2011ApJ...738...79X} samples, and the dots are our samples. In each panel, different colors represent different $\lvert b \rvert $ intervals, with darker colors indicating higher $\lvert b \rvert $.}
	\label{Fig:cp}
\end{figure}

In this way, completeness and purity within each $\lvert b \rvert $ $\&$ $(g-i)_{0}$ intervals are presented in Table~\ref{tab:B1} (see Appendix~\ref{sec:appenb}).  Figure 3 shows the completeness and purity as a function of the stellar color $(g-i)_{0}$. From Figure~\ref{Fig:cp} we can see that the completeness and purity of our classified BHB stars all exceed 90\% for the majority of $\lvert b \rvert $ $\&$ $(g-i)_{0}$ intervals. Moreover, the log $\text{\sl g}$ of BHB stars gradually increases with bluer colors, making them increasingly difficult to distinguish from BS stars. This results in a decrease in
$C$ and $P$ values at bluer colors. On the other hand, both $C$ and $P$ gradually decrease with decreasing $\lvert b \rvert $ due to the increasing reddening at lower latitudes, which consequently leads to larger correction errors in the reddening. 

As an external test, Figure~\ref{Fig:cp} also presents the completeness and purity of the sample from \citet{2011ApJ...738...79X}, represented by hollow symbols. Due to the limited number of available stars, they are grouped based solely on their color. The average completeness and purity of this sample are 88.2\% and 97.6\%, respectively, showing strong consistency with our internal examination.
We note that completeness declines significantly in the reddest bin. This drop is primarily due to the misclassifications of many BHB stars as BS stars by \citet{2011ApJ...738...79X}, as discussed in Section~\ref{sec:2.2}.
Overall, the completeness and purity of the \citet{2011ApJ...738...79X} sample are comparable to those of our sample at $|b| = 20^\circ$. However, the median $|b|$ of the \citet{2011ApJ...738...79X} sample is slightly higher, exceeding $40^\circ$. This moderate discrepancy may be at least partially attributed to unavoidable contamination in the BHB sample selected from spectroscopy.

\section{Distance Calibration}\label{sec:gc}

Calibrating the absolute magnitude ($M_{\rm g}$) of BHB stars is crucial for precisely calculating their distances. Globular clusters (GCs) are usually employed as ideal tracers of distance and metallicity due to the fact that their member stars are expected to born simultaneously from a single giant molecular cloud. Consequently, the stars within GCs exhibit nearly identical distances and chemical patterns. Recently, \cite{2021MNRAS.505.5957B} derived accurate distances for Galactic GCs by mainly using data from HST and Gaia. The BHB stars of GCs are therefore selected in this study for investigating the relations between absolute magnitudes and metallicity as well as color $(g-i)_{0}$. After that, we will use this relation to determine the distances for all BHB sample stars.\\

\begin{figure}
	% \epsscale{1.0}	
	\includegraphics[width=0.5\textwidth]{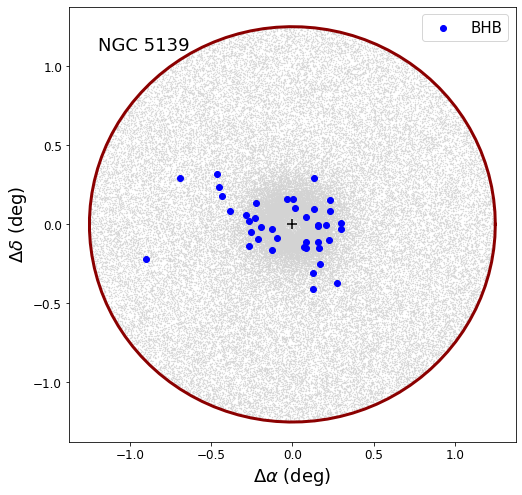}   
	\includegraphics[width=0.5\textwidth]{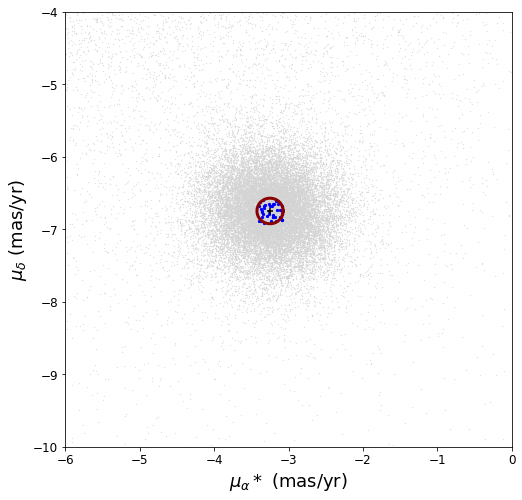}
	\caption{Top panel: BHB stars in NGC 5139 ($\Delta \alpha$, $\Delta \delta$,  represent the relative positions of these stars to the cluster center taken from \cite{1996AJ....112.1487H}, 2010 revision). The black cross marks the center coordinates of the NGC 5139 and the brown ring denotes its 15$r_h$ ($r_h$ is the half-light radius of GC). Bottom panel: proper motion distribution of BHB stars in NGC 5139. The black cross marks the $\mu_{\alpha}* $ and $\mu_{\delta}$ of the NGC 5139 and the brown ring denotes its 8 sigma in proper motion. The gray dots represent the distribution of field stars.}
	\label{Fig:4}
\end{figure}

Here, by using the same methods from adopted from \cite{2020ApJS..249...29H} and \cite{2023ApJ...944...88L}, we select BHB stars located within GCs.  First, the position of BHB stars must within 15 $r_h$ ($r_h$ is the half-light radius of GC) from the center of the GC (from \citealt{1996AJ....112.1487H}, 2010 revision). Subsequently, we select the BHB stars that satisfy the criteria: $\lvert \mu_{\alpha} - \mu_{\rm \alpha,GC} \rvert$ $<$ $8\sigma_{\mu_{\rm \alpha,GC}}$ and $\lvert \mu_{\delta} - \mu_{\rm \delta,GC} \rvert$ $<$ $8\sigma_{\mu_{\rm \delta,GC}}$~, where the proper motions and uncertainties for each GC are taken from \cite{2021MNRAS.505.5978V} \footnote {The adopted $8\sigma_{\mu_{,GC}}$ selection criterion accounts for both the individual stellar proper motion uncertainties and the internal motions of the cluster projected along the tangential direction.} Finally, in this way, 375 BHB stars are selected from 26 GCs. Details regarding these GCs are provided in Table~\ref{tab:gc}. As an example, Figure\,\ref{Fig:4} shows the spatial position and proper motion of our BHB stars (blue dots) in NGC 5139.\\

\begin{table*}[h] 
\setlength\tabcolsep{6pt}
	\setlength{\abovecaptionskip}{0.05cm}
	\centering
	\scriptsize
	\caption{List of all GCs with newly identified BHB.} \label{tab:gc}

	\begin{threeparttable}
		\begin{tabular*}{\hsize}{@{\extracolsep{\fill}} ccrrrrrrr} 
			\hline
			\hline
			Cluster&N ~\tnote{a}& RA (deg) & DEC (deg) & pmRA (mas yr$^{-1}$)~\tnote{b} & pmDEC (mas yr$^{-1}$) ~\tnote{b}& r$_h$ (arcmin)~\tnote{c}& [Fe/H]~\tnote{c} & dis (pc)~\tnote{c} \\
   \hline
  
NGC\ 288 & 22 & 13.189 & $-$26.583 & 4.16 & $-$5.71 & 2.23 & $-$1.32 & 8990\\
  NGC\ 1851 & 3 & 78.528 & $-$40.047 & 2.15 & $-$0.65 & 0.51 & $-$1.18 & 11950\\
  NGC\ 1904 & 4 & 81.046 & $-$24.525 & 2.47 & $-$1.59 & 0.65 & $-$1.6 & 13080\\
  NGC\ 3201 & 16 & 154.403 & $-$46.412 & 8.35 & $-$1.96 & 3.1 & $-$1.59 & 4740\\
  NGC\ 4372 & 2 & 186.439 & $-$72.659 & $-$6.41 & 3.3 & 3.91 & $-$2.17 & 5710\\
  NGC\ 4590 & 14 & 189.867 & $-$26.744 & $-$2.74 & 1.78 & 1.51 & $-$2.23 & 10400\\
  NGC\ 5139 & 40 & 201.697 & $-$47.48 & $-$3.25 & $-$6.75 & 5.0 & $-$1.53 & 5430\\
  NGC\ 5272 & 38 & 205.548 & 28.377 & $-$0.15 & $-$2.67 & 2.31 & $-$1.5 & 10180\\
  NGC\ 5904 & 26 & 229.638 & 2.081 & 4.09 & $-$9.87 & 1.77 & $-$1.29 & 7480\\
  NGC\ 6121 & 4 & 245.897 & $-$26.526 & $-$12.51 & $-$19.02 & 4.33 & $-$1.16 & 1850\\
  NGC\ 6205 & 33 & 250.422 & 36.46 & $-$3.15 & $-$2.57 & 1.69 & $-$1.53 & 7420\\
  NGC\ 6218 & 25 & 251.809 & $-$1.949 & $-$0.19 & $-$6.8 & 1.77 & $-$1.37 & 5110\\
  NGC\ 6254 & 7 & 254.288 & $-$4.1 & $-$4.76 & $-$6.6 & 1.95 & $-$1.56 & 5070\\
  NGC\ 6341 & 28 & 259.281 & 43.136 & $-$4.94 & $-$0.63 & 1.02 & $-$2.31 & 8500\\
  NGC\ 6362 & 15 & 262.979 & $-$67.048 & $-$5.51 & $-$4.76 & 2.05 & $-$0.99 & 7650\\
  NGC\ 6397 & 7 & 265.175 & $-$53.674 & 3.26 & $-$17.66 & 2.9 & $-$2.02 & 2480\\
  NGC\ 6541 & 14 & 272.01 & $-$43.715 & 0.29 & $-$8.85 & 1.06 & $-$1.81 & 7610\\
  NGC\ 6752 & 6 & 287.717 & $-$59.985 & $-$3.16 & $-$4.03 & 1.91 & $-$1.54 & 4120\\
  NGC\ 6809 & 17 & 294.999 & $-$30.965 & $-$3.43 & $-$9.31 & 2.83 & $-$1.94 & 5350\\
  NGC\ 7078 & 12 & 322.493 & 12.167 & $-$0.66 & $-$3.8 & 1.0 & $-$2.37 & 10710\\
  NGC\ 7089 & 8 & 323.363 & $-$0.823 & 3.44 & $-$2.16 & 1.06 & $-$1.65 & 11690\\
  NGC\ 7099 & 34 & 325.092 & $-$23.18 & $-$0.74 & $-$7.3 & 1.03 & $-$2.27 & 8460\\
\hline
			\hline
			
		\end{tabular*}
		\begin{tablenotes}
			\footnotesize
                \item[a] The number of BHB stars in GC.
			\item[b] \cite{2021MNRAS.505.5978V}.
			\item[c] \cite{1996AJ....112.1487H}.
		\end{tablenotes}	
	\end{threeparttable}
\end{table*}

\begin{figure}[htbp] 
	\centering
	\includegraphics[width=0.5\textwidth]{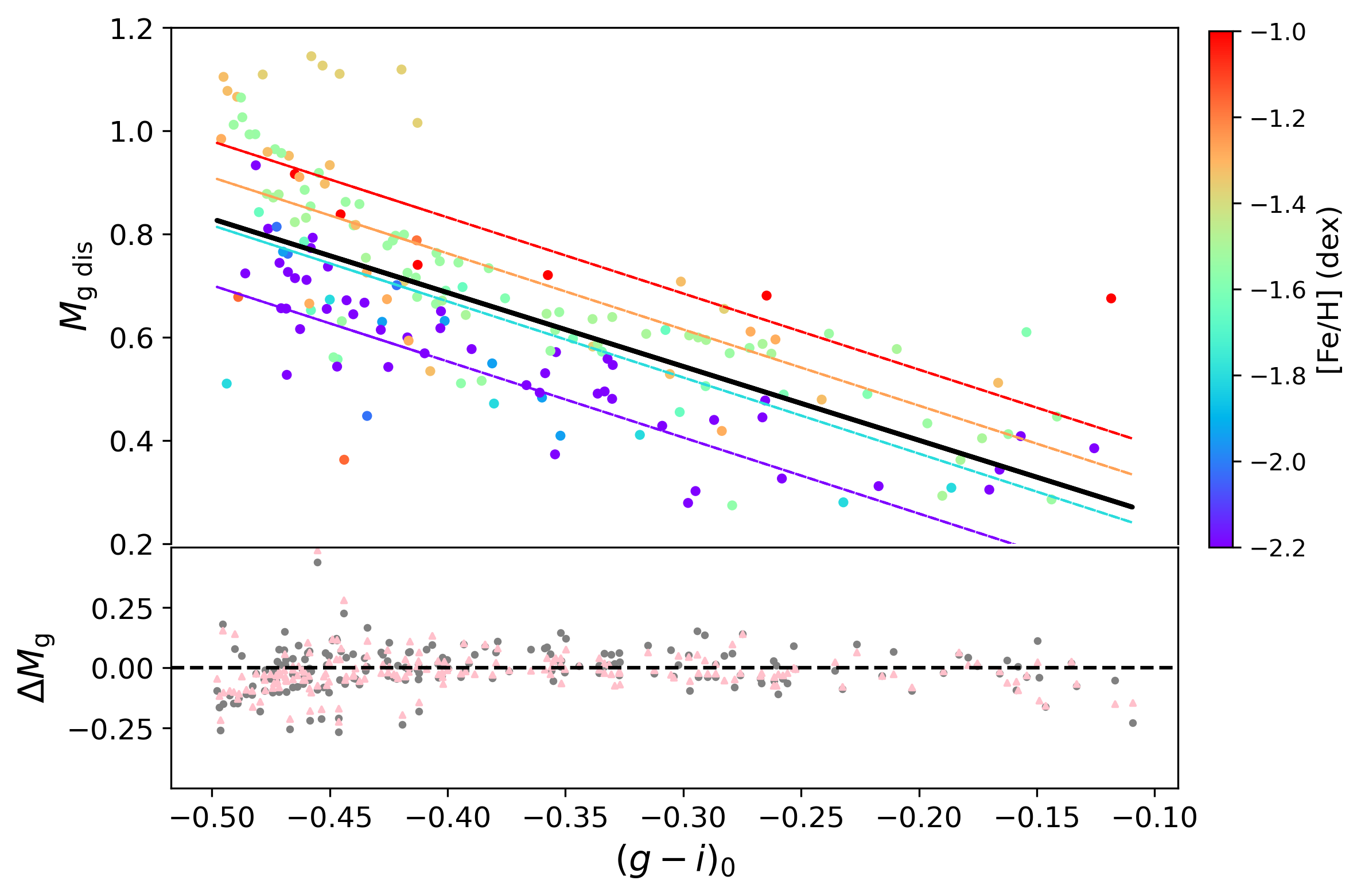} 
	\caption{Top panel: the distribution of half BHB stars in GCs in the $M_{g}^{\rm dis}$ vs $(g-i)_{0}$ plane with a color-coded [Fe/H] (The [Fe/H] is sourced from \citealt{1996AJ....112.1487H}). The colored lines are 4 different [Fe/H] fitted to the points using Euqation (3). The black line are linear model fitted to the points (Equation 4).  Bottom panel: the residuals of the calculated first-order 2D polynomial fitting results (pink triangles) and 
    linear model fitting results (grey dots).}
	\label{Fig:5}
\end{figure}

Half of the GC BHB member stars (hereafter training sample) are adopted to explore the relation between absolute magnitudes and stellar color as well as the metallicity, while the remaining half BHB stars (hereafter testing sample) are used to check the calibrated relation. First, we calculate $g$-band absolute magnitudes of all GC BHB member stars by using the distances of their corresponding GCs ($M_{g}^{\rm dis}$). As mentioned, the distances of GCs are all taken from \cite{2021MNRAS.505.5957B}. The extinction corrections are using reddening values from \cite{1998ApJ...500..525S}. The relation between $M_{g}^{\rm dis}$ and $(g-i)_{0}$ for the training sample is displayed in Figure\,~\ref{Fig:5}. The metallicity information of these GC BHB members is taken from \cite{1996AJ....112.1487H}. From this plot, it can be seen that, at the same $(g-i)_{0}$, more metal-rich stars exhibit larger $M_{g}^{\rm dis}$ (lower luminosity). At the same metallcity, as the stellar color becomes redder, $M_{g}^{\rm dis}$ decreases (higher luminosity). We apply a first-order 2D polynomial fitting to the absolute magnitude for BHB stars in GCs, as a function of their $(g-i)_{0}$, and [Fe/H]:

\begin{equation}
M_{\rm g} = a_0 \times (g-i)_{0} + a_1 \times \rm [Fe/H] + a_2.\\
\end{equation} 

The parameters ($a_0$, $a_1$, $a_2$) are fitted by maximising the likelihood function, employing the Markov Chain Monte Carlo (MCMC) ensemble sampler \texttt{emcee} \citep{2013PASP..125..306F} to sample the posterior distribution. The best-fit coefficients are $-$1.43$\pm$0.004, 0.23$\pm$0.005 and 0.49$\pm$0.010 for $a_0$, $a_1$ and $a_2$, respectively. The colored lines in top panel of Figure\,~\ref{Fig:5} represent our best fits for four different [Fe/H] with values ranging from $-$1.0\,dex (red) to $-$2.2\,dex (purple), shown as examples. The pink triangles in bottom panel of Figure\,~\ref{Fig:5} shows the residuals (the difference between $M_g^{\rm dis}$ and these predicted by Equation 3). The standard of the residuals is only 0.07\,mag, corresponding to a distance precision as well as 3.2\%. However, estimating the metallicity of the BHB stars identified in this work is difficult without spectroscopic data. As a result, we focus on deriving a relationship between the
$g$-band absolute magnitude and stellar color $(g-i)_{0}$ alone. A linear function is adopted to describe this relation: 

\begin{equation}
M_{\rm g} = A \times (g-i)_{0} + B. \\
\end{equation}

\begin{figure}[htbp] 
	\centering
	\includegraphics[width=0.5\textwidth]{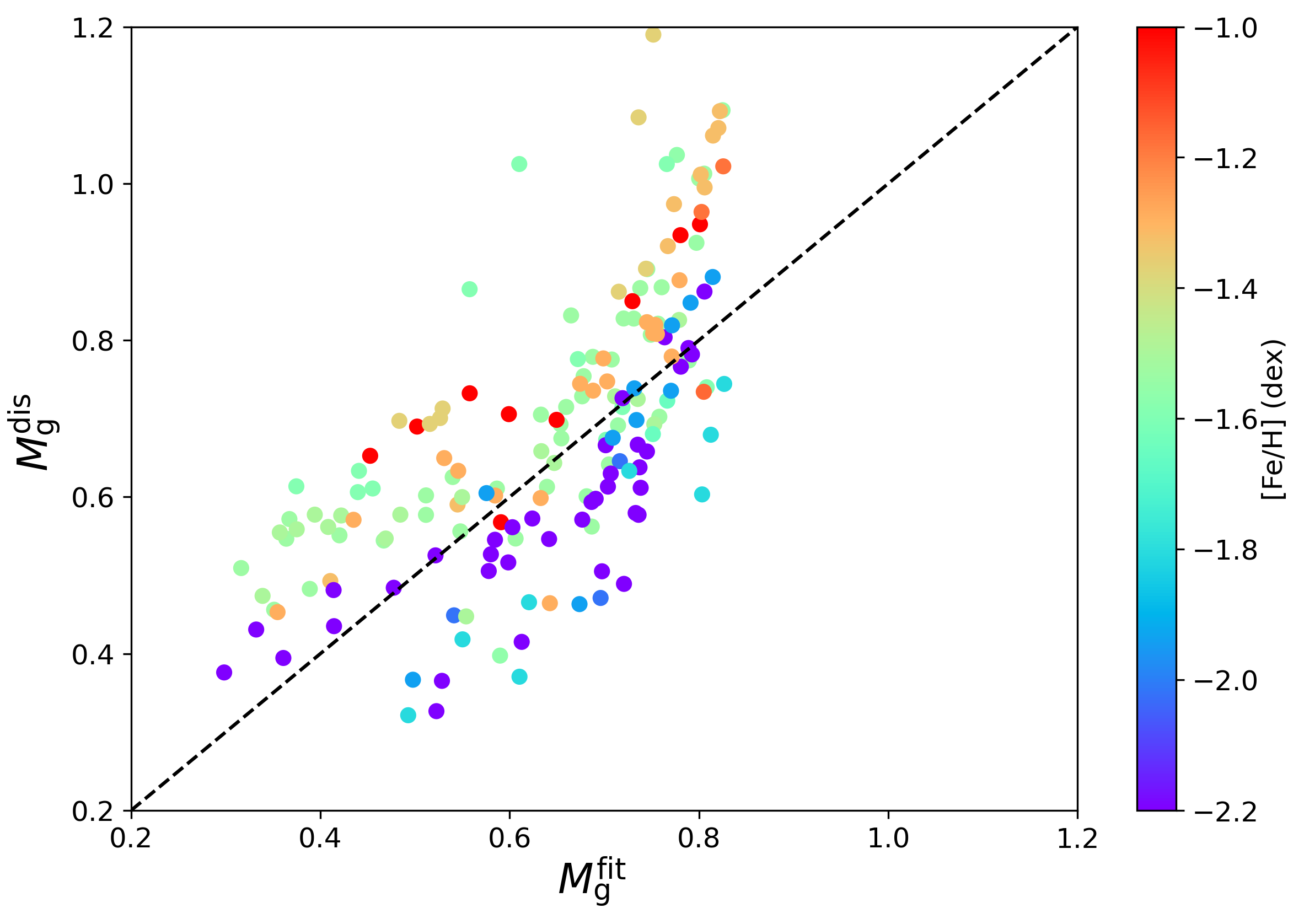} 
	\caption{The comparison of $M_{g}^{\rm dis}$ obtained using GC distances and $M_{g}^{\rm fit}$ which obtained by the relationship between $(g-i)_{0}$ and $g$-band absolute magnitude (Equation 4).}
	\label{Fig:6}
\end{figure}

Again using MCMC, the best-fit of the parameters are $A$= $-$1.390 $\pm$ 0.009, and $B$ = 0.132 $\pm$ 0.004 (represented by black line in Figure\,~\ref{Fig:5}). In the bottom panel of Figure\,~\ref{Fig:5}, grey dots shows the residuals, the standard of the residuals is 0.11\,mag, corresponding to a distance precision as well as 5\%. It should be noticed that the best fit line pass through stars at [Fe/H] $<$ -1.5 dex from Figure\,~\ref{Fig:5}, if a star is more metal rich, the distance would be overestimated. To independently verify the accuracy of the derived calibrated relation, we compare the predicted $g$-band absolute magnitudes from this relation with those obtained using the cluster distances of BHB stars in the testing sample. The results are presented in Figure\,~\ref{Fig:6}. Overall, the scatter is approximately 0.11\,mag, consistent with that of the training sample. However, we note a potential systematic offset ranging from 0 to 0.05\,mag, which could arise from the neglect of metallicity effects. We emphasize that the metallicity-dependent calibration can be applied to spectroscopic samples to derive their distances with exceptional precision, achieving an accuracy within 3\%.\\

\begin{figure}[htbp] 
	\centering
	\includegraphics[width=0.5\textwidth]{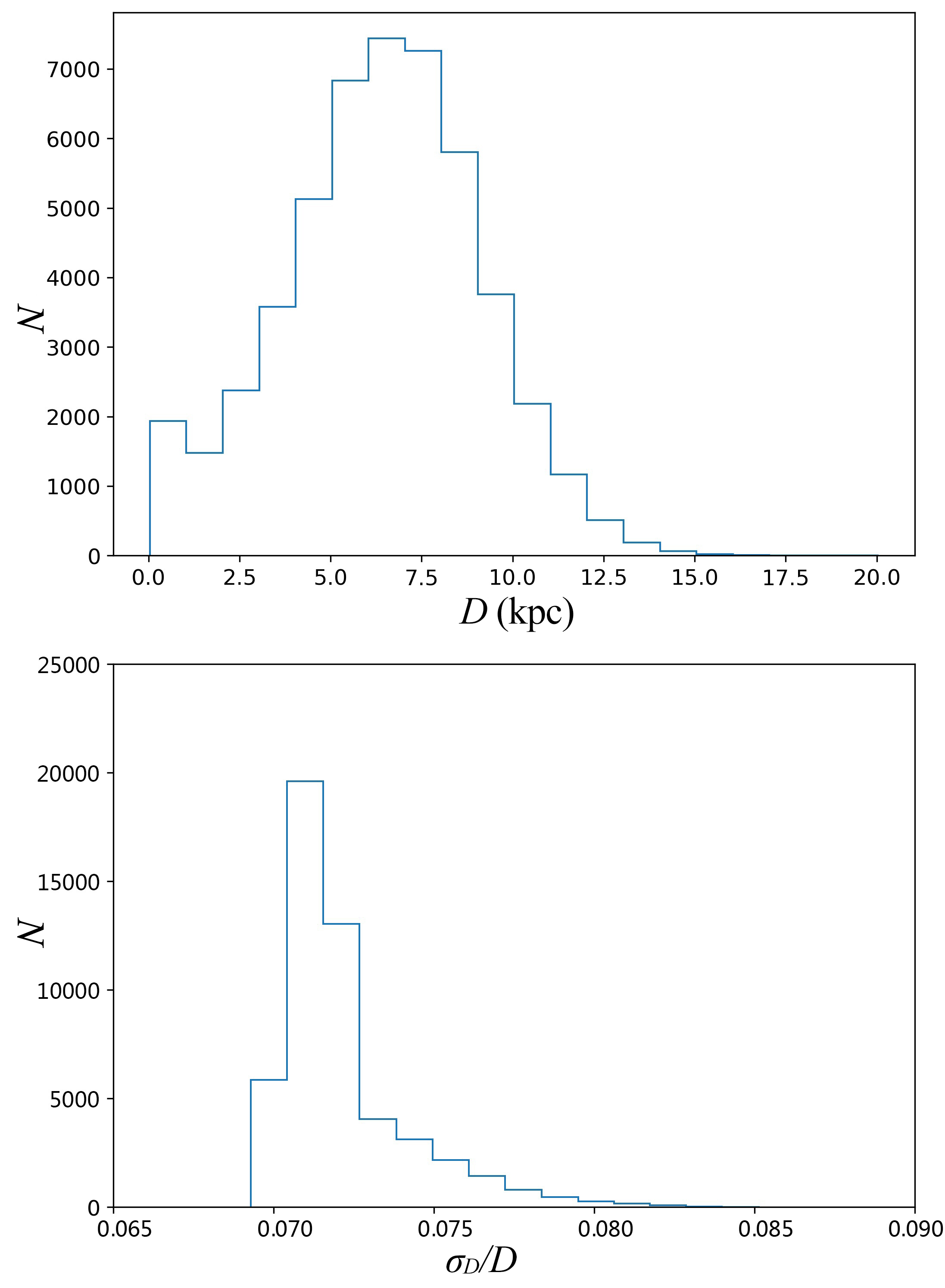} 
	\caption{Top panel: the distribution of the distances of our BHB samples. Bottom panel: the distribution of the relative distance errors.}
	\label{Fig:dis}
\end{figure}

By applying the above relationship, we obtain the $M_{\rm g}$ for all BHB stars and then calculate their distances. The distribution of distances for these BHB stars are shown in Figure\,~\ref{Fig:dis}. The most distant BHB stars can reach as far as 20 kpc. Furthermore, the distribution of relative distance errors is given in the bottom panel of Figure\,~\ref{Fig:dis}, with the typical relative precision of distance estimated at 7.1$\%$. The precise distances of BHB stars serve as a robust foundation for further exploration of the Galactic halo.
  
\section{Discussion} 

\subsection{Spatial Distribution of Our BHB stars}
\begin{figure}[htbp] 
	\centering
	\includegraphics[width=0.5\textwidth]{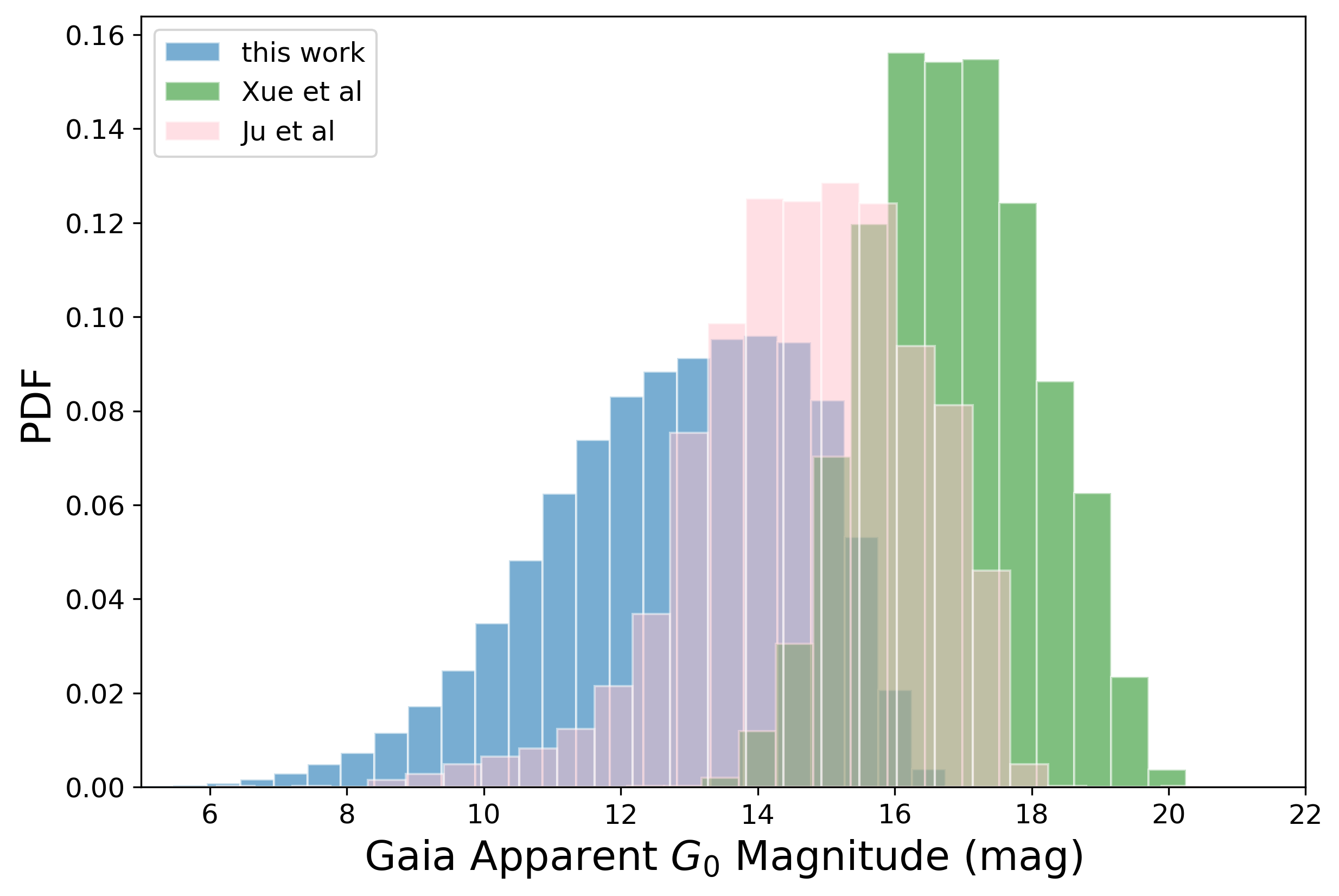} 
	\caption{The distribution of apparent magnitudes for BHB of this work (blue), \citeauthor{2011ApJ...738...79X} (\citeyear{2011ApJ...738...79X}; green), and \citeauthor{2024ApJS..270...11J} (\citeyear{2024ApJS..270...11J}; pink). }
	\label{Fig:gmag}
\end{figure}

BHB stars are generally regarded as possessing a nearly constant absolute magnitudes. Consequently, their apparent magnitude can serve as a reliable distance indicator, in other words, farther BHB stars tend to have larger apparent magnitudes. We compare the $G$ magnitude of our BHB stars with those from \cite{2011ApJ...738...79X} and \cite{2024ApJS..270...11J}, as depicted in Figure\,~\ref{Fig:gmag}, where extinction corrections are made using the reddening maps from \cite{1998ApJ...500..525S}. The peak $G$ magnitudes of our BHB sample, as well as those from \cite{2011ApJ...738...79X} and \cite{2024ApJS..270...11J}, are approximately 13.0\,mag, 14.6\,mag, and 16.8\,mag, respectively. Compared to these two works, our BHB samples appear brighter, suggesting they are located closer. \\

\begin{figure}[htbp]
\centering
\begin{minipage}{.5\textwidth}
  \centering
  \includegraphics[width=.8\linewidth]{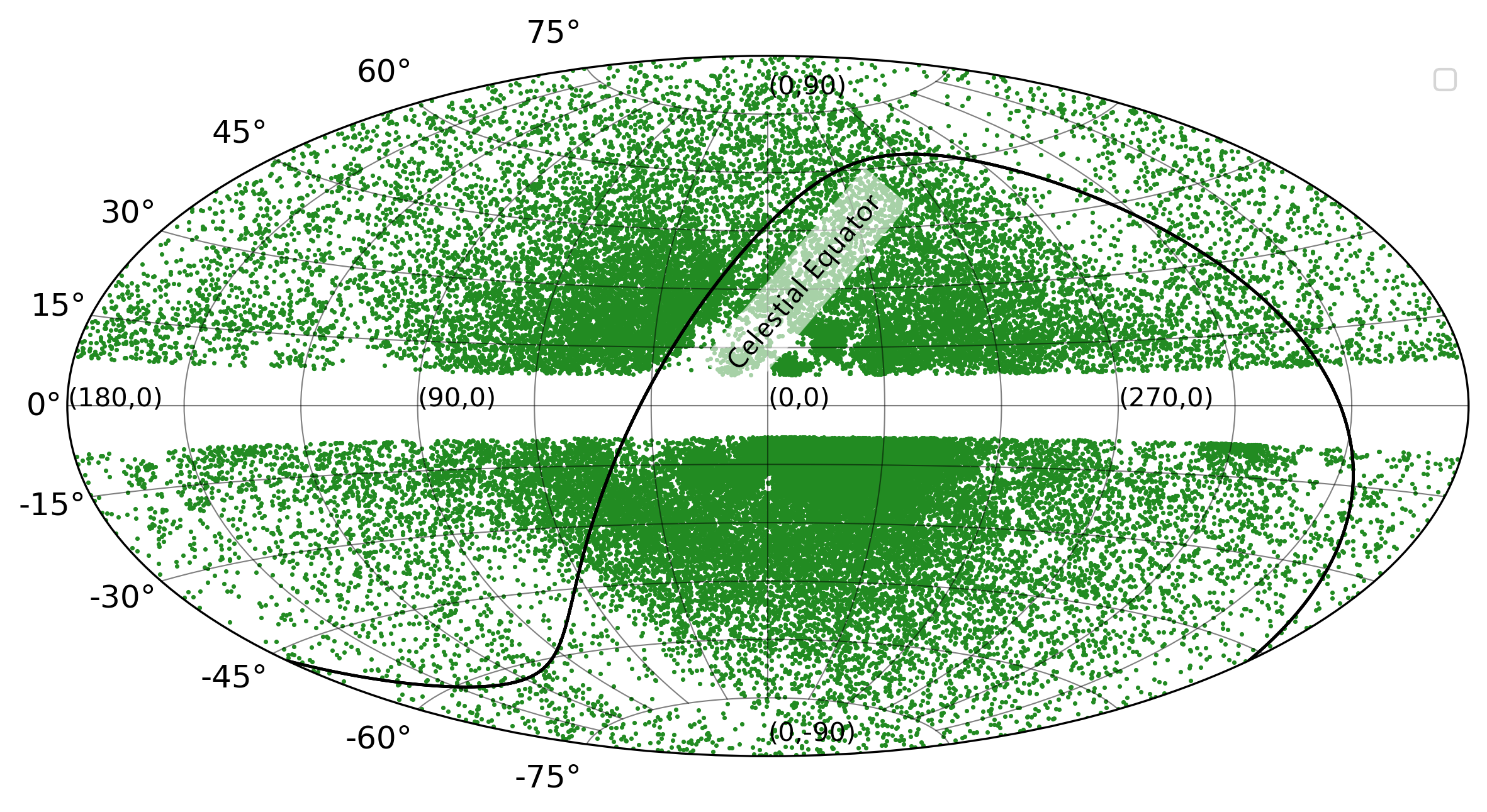}  
  \label{fig:fig1}
\end{minipage}%
\begin{minipage}{.5\textwidth}
  \centering
  \includegraphics[width=.8\linewidth]{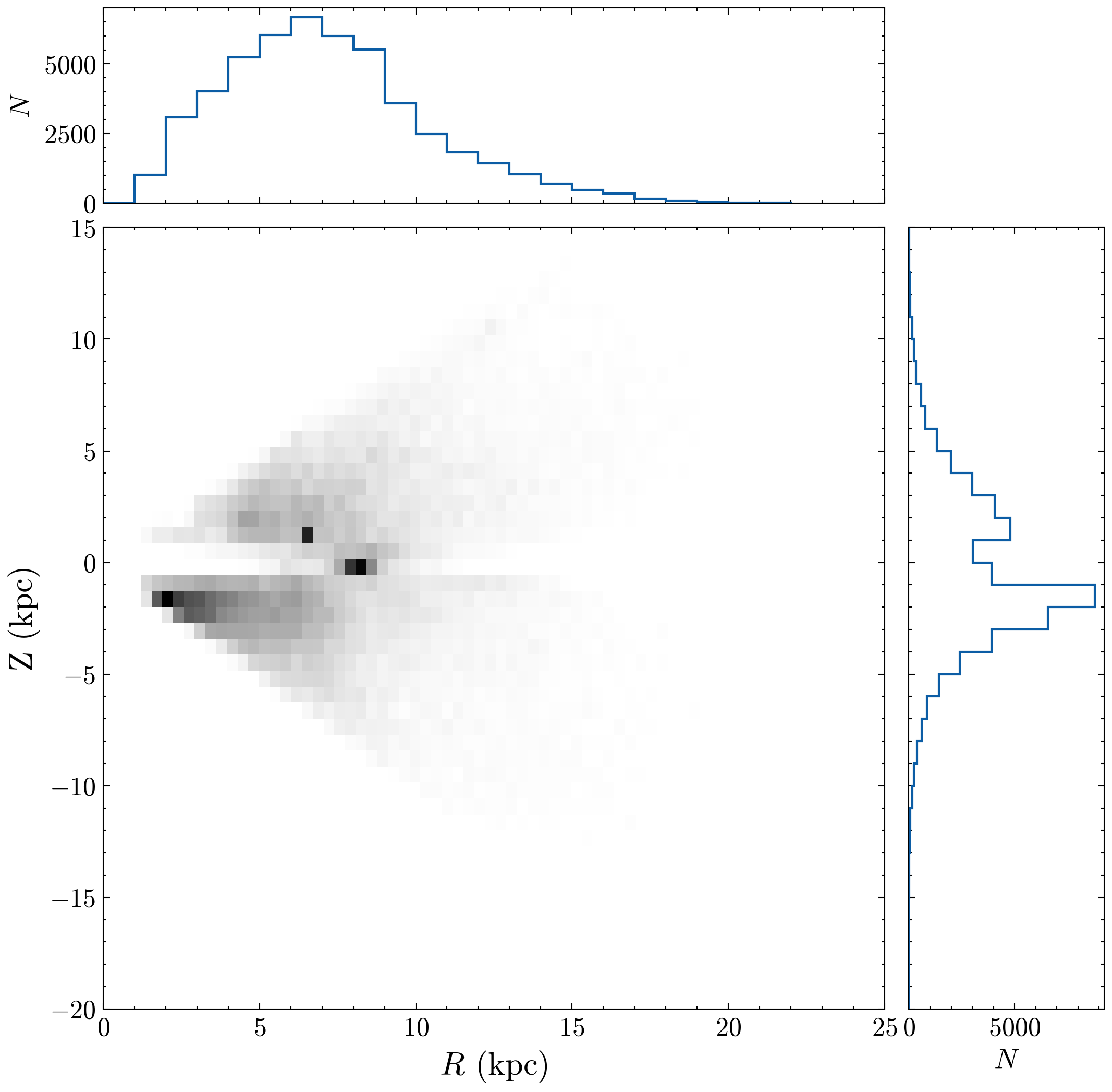}
  \caption{Top panel: the Galactic sky coverage of our BHB sample in celestial equator coordinate system; Bottom panel: the spatial distribution of our BHB samples in the $R$ - $Z$ plane.}
  \label{fig:spatial}
\end{minipage}
\end{figure}

The Galactic sky coverage and spatial distribution of our BHB stars are illustrated in Figure\,~\ref{fig:spatial}, showing that most of our BHB stars are located within $\lvert Z \rvert $ $<$ 10 kpc. Furthermore, the right histogram reveals that the peak value of the $\lvert Z \rvert $ is about 3 kpc, indicating that most of our BHB stars are located in the inner halo. Notably, our sample includes many BHB stars with low Z values, inspiring us to explore the origins of these stars. Furthermore,  our sample encompasses a volume closer to the Sun than previous known samples, providing valuable data for studying the properties of the Galactic halo, particularly in our solar neighborhood. 

% Combining our samples with \cite{2011ApJ...738...79X} and \cite{2024ApJS..270...11J} samples, we could map the Galatic halo from the inner to the outer.

\subsection{On the performance identification BHB from CSST}
In the future, the CSST, a 2 m space telescope with a field of view 1.1\,deg$^2$ (\citealt{2011SSPMA..41.1441Z,2018cosp...42E3821Z}) is planned to be launched in 2027 to carry out large-scale sky survey covering nearly 17,500\,deg$^2$ at high Galactic latitude (b $\geq$ 15$^\circ$) with $g$ band limiting magnitude down to around 26.3 (5$\sigma$ point-like source). The wide-field multiband large-scale sky survey of CSST,  will cover a wavelength ranging from 2000$\overset{\circ}{\rm A}$ to 1.1 $\mu$m, which is divided into seven bands (near-ultraviolet ($NUV$), $u, g, r, i, z, y$). Given its such large sky coverage and survey depth, the forthcoming CSST data will offer a groundbreaking perspective of our Galaxy, particularly unveiling unprecedented insights into the outer halo within a significant volume (\citealt{2023SCPMA..6619511C}). In this section, we explore the capability of selecting BHBs from CSST broad-band photometry.\\

\begin{figure}[htbp] 
	\centering
	\includegraphics[width=0.5\textwidth]{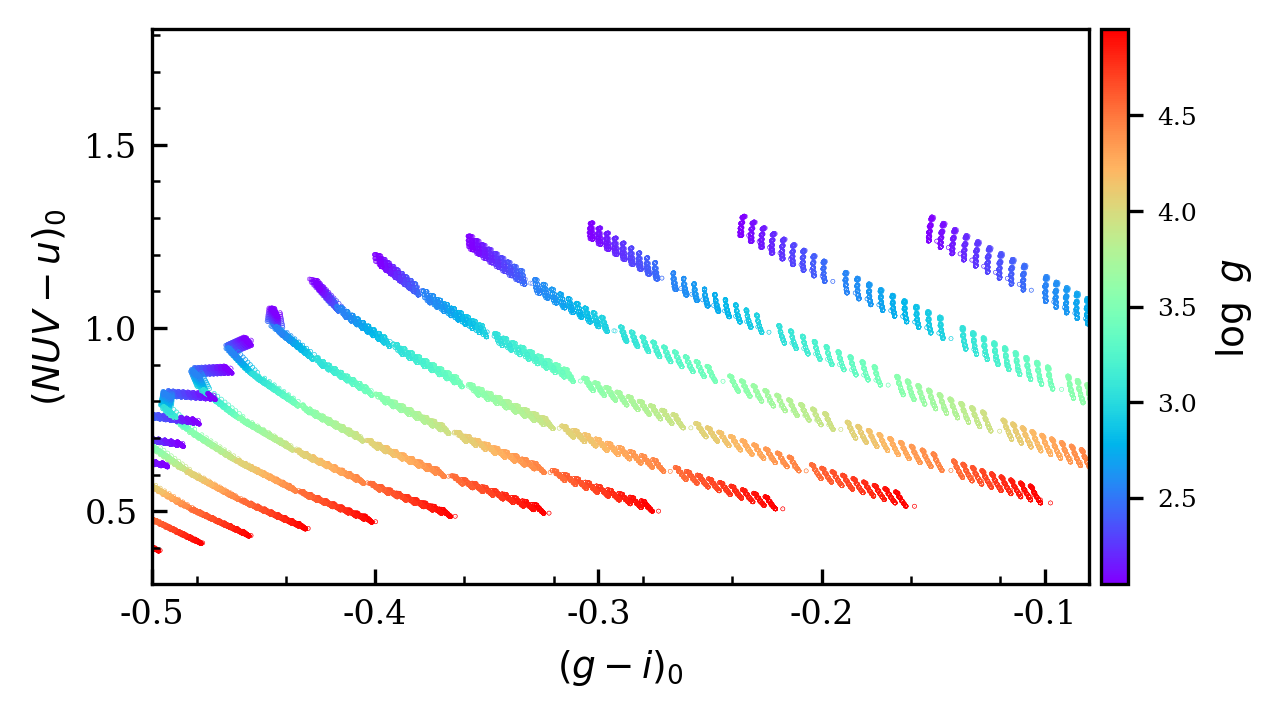} 
	\caption{Sensitivity of color to log $\text{\sl g}$ of stars from CSST photometry in the $(NUV - u)_0$ versus $(g - i)_0$ plane, color coded by log $\text{\sl g}$, as shown by the right color bar. The values of log $\text{\sl g}$ range from 2\, to 5\,dex.}
	\label{Fig:csst1}
\end{figure}

To calculate the synthetic magnitudes, convolve theoretical spectra with the transmission curves of the filter systems of CSST \citep{2014MNRAS.444..392C}. Using the methodology outlined by \cite{2012PASP..124..140B}, we determine the synthetic magnitude in the AB system \citep{1983ApJ...266..713O,1996AJ....111.1748F}, referencing Equations 1 to 3 in \cite{2024RAA....24d5015S}. In this work, we adopt the spectra from the BOSZ synthetic library ($R$ $\sim$ 2000) \footnote{https://archive.stsci.edu/prepds/bosz/} \citep{2017AJ....153..234B}. For a detailed parameter coverages ($T$$\rm _{eff}$, log $\text{\sl g}$, [Fe/H], etc.), we direct the reader to \cite{2017AJ....153..234B}.\\

Color $(NUV - u)$ is thought as an indicator of surface gravity, and has been validated by \cite{2024RAA....24d5015S}. Taking this into account,  we assess the classification of BHB and BS stars using this color. Considering the original parameter of log $\text{\sl g}$ space coverage is too sparse (0.5\,dex), we thus interpolate the BOSZ theoretical spectra to a step of 0.05\,dex for 2 $<$ log $\text{\sl g}$ $<$ 5\,dex. In the newly generated log $\text{\sl g}$, the $``SpecificIntensity"$ corresponding to the wavelength is estimated through linear interpolation of the $``SpecificIntensity"$ located in the ``shoulder" region on either side of the log $\text{\sl g}$.  The result is shown in Figure\,~\ref{Fig:csst1}, and we can see that $(NUV - u)_0$ shows significant sensitivity to log $\text{\sl g}$ between $(g - i)_0$ $>$ $-$0.5 and $(g - i)_0$ $<$ 0.08, which includes the color ranges of both BHB and BS stars.\\

\begin{figure}[b] 
	\centering
	\includegraphics[width=0.5\textwidth]{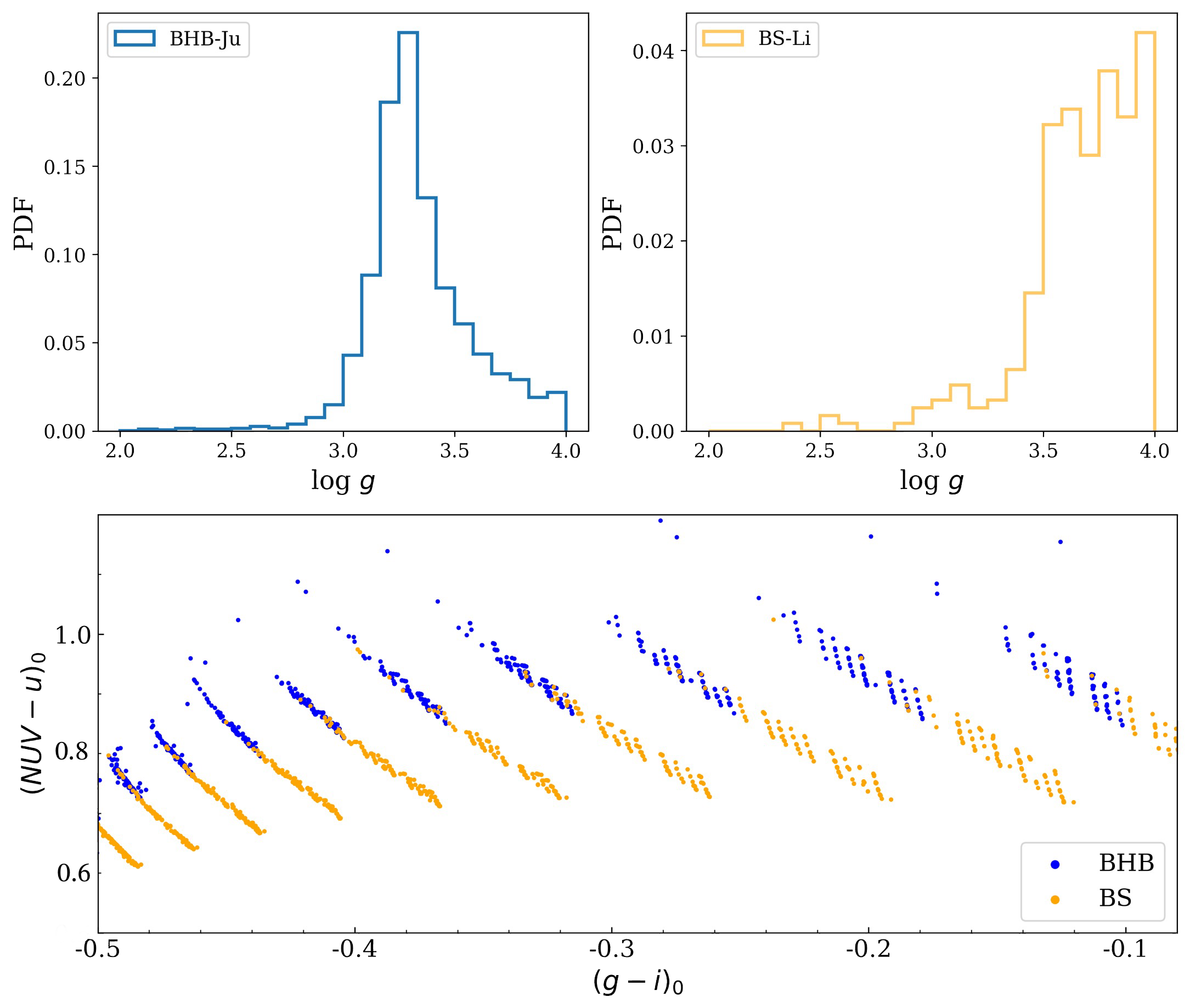} 
	\caption{Top left panel: the log $\text{\sl g}$ distribution of BHB. Top right panel: the log $\text{\sl g}$ distribution of BS. Bottom panel: simulations of the real distributions of BHB and BS stars in $(NUV - u)_0$ versus $(g - i)_0$ plane.}
	\label{Fig:csst2}
\end{figure}

To simulate a realistic situation, we utilize the observational surface gravity distributions of BHB and BS stars from the samples in \cite{2024ApJS..270...11J} and \cite{2023A&A...672A..81L}, respectively. The log $\text{\sl g}$ distribution of BHB and BS stars (shown in the top panels of Figure\,~\ref{Fig:csst2}) are employed to re-sample the number of used theoretical spectra and their colors. The result is shown in the bottom panel of Figure\,~\ref{Fig:csst2}. Notably, we observe distinct sequences in the 
$(NUV - u)_0$ versus $(g - i)_0$ distribution of BHBs (depicted by blue dots) and BS (represented by orange dots) stars. Therefore, according to our mock data tests, distinguishing between BHB and BS stars for CSST is feasible. Using similar double Gaussian fitting technique established in Section\,~\ref{sec:2.2}, the predicted completeness and purity of our results from the mock data tests are 88.73\% and 76.64\%, respectively, if assuming a typical uncertainty of 0.01\,mag for both  $(NUV - u)$ and $(g - i)$ colors and an extinction correction error of 0.005\,mag in both colors. Given the limiting magnitude of CSST (\(r = 23\,\mathrm{mag}\)) and the absolute magnitude of BHB stars, the BHB sample identified by CSST in the future will enable a more detailed exploration of the outer halo of our Galaxy, reaching distances of up to approximately \(300\,\mathrm{kpc}\).\\

\section{Summary}

 In this work, we successfully identify 49,733 BHB stars based on synthetic colors $(u-v)_{0}$ and $(g-i)_{0}$ in SkyMapper photometric systems, which are convolved from Gaia XP spectra. They are identified as BHB stars with completeness and purity exceeding 90\% in $|b| \geq 8^{\circ}$. \\

 We investigate the relationship between absolute magnitudes and color using globular clusters as calibrators. We determine the distances of our BHB stars by employing this relationship. The distance distribution is broad, extending beyond 20 kpc, with a typical relative error of 7.1$\%$. The accurate measurement of BHB distances enables us to better study the kinematics and structure of the Milky Way. Furthermore, the majority of our BHB sample stars are located in the inner halo. \\
 
Using the BOSZ theoretical spectral library and the observational log $\text{\sl g}$ distribution of BHB and BS stars, we generate mock data to simulate the real distributions of BHB and BS stars in the CSST $(NUV - u)_0$ versus $(g - i)_0$  plane. According to our mock data sets,  BHB stars can be distinguished from BS stars efficiently by using upcoming photometry from CSST. It is inspiring that the structure of the Milky Way can be further revealed using such a large sample of BHB stars from CSST. The large sample of BHB stars from CSST will significantly improve our understanding of the Milky Way structure, particularly in the outer halo, where previous studies have been limited by smaller sample sizes. This will not only allow us to trace the Galactic halo's three-dimensional shape with unprecedented precision, but also provide critical constraints on the Milky Way's assembly history and the distribution of dark matter.\\

\vspace{7mm} \noindent {\bf Acknowledgments}

This study is supported by the National Natural Science Foundation of China under grant No. 12173013; the Postdoctoral Fellowship Program of CPSF under Grant Number GZC20240371; the project of Hebei Provincial Department of Science and Technology under grant number 226Z7604G; Science Foundation of Hebei Normal University (Nos. L2024B55, L2024B54, and L2024B56) and the Hebei NSF (No. A2021205006). We acknowledge the China Manned Space Program with grant no. CMS-CSST-2025-A13. \\

The national facility capability for SkyMapper has been funded through ARC LIEF grant LE130100104 from the Australian Research Council, awarded to the University of Sydney, the Australian National University, Swinburne University of Technology, the University of Queensland, the University of Western Australia, the University of Melbourne, Curtin University of Technology, Monash University and the Australian Astronomical Observatory. SkyMapper is owned and operated by The Australian National University's Research School of Astronomy and Astrophysics. The survey data were processed and provided by the SkyMapper Team at ANU. The SkyMapper node of the All-Sky Virtual Observatory (ASVO) is hosted at the National Computational Infrastructure (NCI). Development and support of the SkyMapper node of the ASVO has been funded in part by Astronomy Australia Limited (AAL) and the Australian Government through the Commonwealth's Education Investment Fund (EIF) and National Collaborative Research Infrastructure Strategy (NCRIS), particularly the National eResearch Collaboration Tools and Resources (NeCTAR) and the Australian National Data Service Projects (ANDS).\\ 

This work has made use of data from the European Space Agency (ESA) mission {\it Gaia} (\url{https://www.cosmos.esa.int/gaia}), processed by the Gaia Data Processing and Analysis Consortium (DPAC; \url{https:// www.cosmos.esa.int/web/gaia/dpac/ consortium}). Funding for the DPAC has been provided by national institutions, in particular, the institutions participating in the Gaia Multilateral Agreement.

\software{Numpy \citep{2011CSE....13b..22V}, Scipy \citep{2007CSE.....9c..10O}, Matplotlib \citep{2007CSE.....9...90H}, \texttt{emcee} \citep{2013PASP..125..306F}.}

\bibliography{ref}{}
\bibliographystyle{aasjournal}

\clearpage

\appendices

\setcounter{table}{0}  
\section{Appendix A: The parameters of BHB stars in different $\lvert b \rvert $ $\&$ $(g-i)_{0}$ intervals.}\label{sec:appenb}
\renewcommand{\thetable}{A\arabic{table}}

\begin{table*}[htbp]
\centering\small
\begin{threeparttable}
\caption{\label{tab:B1}The parameters of BHB stars in different $\lvert b \rvert $ $\&$ $(g-i)_{0}$ intervals}
\begin{tabular}{cccccc}
\hline
  \makebox[0.1\textwidth][c]{$\lvert b \rvert $ }& \makebox[0.12\textwidth][c]{$(g-i)_{0}$}&  \makebox[0.12\textwidth][c]{$N_{\rm BHB}$~\tnote{a}} &  \makebox[0.12\textwidth][c]{$Pro_{\rm BHB}$~\tnote{b}} & \makebox[0.12\textwidth][c]{$C$ ~\tnote{c}}&  \makebox[0.12\textwidth][c]{$P$ ~\tnote{d}}\\ \midrule
&  [-0.50, -0.45] & 355 &   0.78  &0.988 & 0.988\\

 &  [-0.45, -0.40] & 370&0.70& 0.993 & 0.988\\

 &  [-0.40, -0.35] & 330&0.64&0.998 & 0.995\\

[65$^\circ$, 90$^\circ$] &  [-0.35, -0.30] & 241&0.52&0.998 & 0.998\\

 &  [-0.30, -0.25] &205&0.42&0.999 & 0.999\\

 &  [-0.25, -0.20] & 154&0.30&0.997 & 0.997\\

 &  [-0.20, -0.08] & 206&0.14&0.998 & 0.998\\
\hline

&  [-0.50, -0.45] & 399&0.75&0.968 & 0.940\\

&  [-0.45, -0.40] & 480&0.73&0.993 & 0.988\\

&  [-0.40, -0.35] & 375&0.66&0.997 & 0.996\\

[55$^\circ$, 65$^\circ$] &  [-0.35, -0.30] &274&0.53& 0.998 & 0.998\\

& [-0.30, -0.25] & 222&0.41&0.998 & 0.997\\

&  [-0.25, -0.20] &199&0.32& 0.998 & 0.998\\

&  [-0.20, -0.08] &213&0.12 &0.998 & 0.998\\
\hline

& [-0.50, -0.45] & 797&0.77&0.981 & 0.971\\

&  [-0.45, -0.40] &716&0.71&0.991 & 0.983\\

&  [-0.40, -0.35] &618&0.65& 0.998 & 0.997\\

[45$^\circ$, 55$^\circ$] &  [-0.35, -0.30] &445&0.51& 0.997 & 0.995\\

&  [-0.30, -0.25] &365&0.40& 0.998 & 0.998\\

& [-0.25, -0.20] & 315&0.29&0.999 & 0.999\\

&  [-0.20, -0.08] & 350&0.12&0.997 & 0.997\\
\hline

&  [-0.50, -0.45] & 1397&0.76&0.971 & 0.965\\

&  [-0.45, -0.40] & 1267&0.69&0.991 & 0.987\\

&  [-0.40, -0.35] &978&0.60& 0.994 & 0.991\\

[35$^\circ$, 45$^\circ$] &  [-0.35, -0.30] & 755&0.47&0.993 & 0.992\\

&  [-0.30, -0.25] & 533&0.33&0.993 & 0.991\\

& [-0.25, -0.20] & 459&0.24&0.997 & 0.996\\

&  [-0.20, -0.08] & 570&0.10&0.999 & 0.990\\
\hline

&  [-0.50, -0.45] & 2161&0.67&0.942 & 0.942\\

&  [-0.45, -0.40] & 2002&0.60&0.982 & 0.973\\

&  [-0.40, -0.35] & 1600&0.49&0.991 & 0.989\\

[25$^\circ$, 35$^\circ$] & [-0.35, -0.30] & 1188&0.37&0.993 & 0.992\\

&  [-0.30, -0.25] &872&0.25& 0.992 & 0.994\\

&  [-0.25, -0.20] &750&0.18& 0.997 & 0.997\\

&  [-0.20, -0.08] &905&0.07& 0.997 & 0.997\\
\hline

&  [-0.50, -0.45] &3674&0.50& 0.878 & 0.899\\

&  [-0.45, -0.40] & 3170&0.40&0.954 & 0.950\\

&  [-0.40, -0.35]  &2377&0.29& 0.977 & 0.974\\

[15$^\circ$, 25$^\circ$] &  [-0.35, -0.30] & 1664&0.19&0.983 & 0.982\\

&  [-0.30, -0.25]& 1276&0.13&0.982 & 0.986\\

&  [-0.25, -0.20] &1087&0.09& 0.977 & 0.986\\

&  [-0.20, -0.08] & 1134&0.03&0.989 & 0.992\\
\hline

&  [-0.50, -0.45] & 2904&0.15&0.693 & 0.767\\

&  [-0.45, -0.40] & 2648&0.12&0.869 & 0.878\\

&  [-0.40, -0.35] & 2040&0.09&0.931 & 0.921\\

[8$^\circ$, 15$^\circ$] &  [-0.35, -0.30] &1465&0.06& 0.951 & 0.958\\

&  [-0.30, -0.25] & 1207&0.04&0.967 & 0.972\\

&  [-0.25, -0.20] & 1001&0.03&0.957 & 0.967\\

&  [-0.20, -0.08] & 1018&0.01&0.931 & 0.959\\

\bottomrule \end{tabular} \small
\begin{tablenotes}
 \item[a] The number of BHB in this $\lvert b \rvert $ $\&$ $(g-i)_{0}$ interval.
 \item[b] The proportion of BHB in this $\lvert b \rvert $ $\&$ $(g-i)_{0}$ interval.
 \item[c] Completeness.
 \item[d] Purity.
\end{tablenotes} 
\end{threeparttable}

\end{table*}

\clearpage

 \section{Appendix B: The distributions of $(u-v)_{0}$ of six supplementary $\lvert b \rvert $ intervals}\label{sec:appena}
\setcounter{figure}{0}
\renewcommand{\thetable}{B\arabic{table}}
\renewcommand{\thefigure}{B\arabic{figure}}

\begin{figure}[b]
  \begin{minipage}[t]{0.5\linewidth}
    \centering
    \includegraphics[scale=0.5]{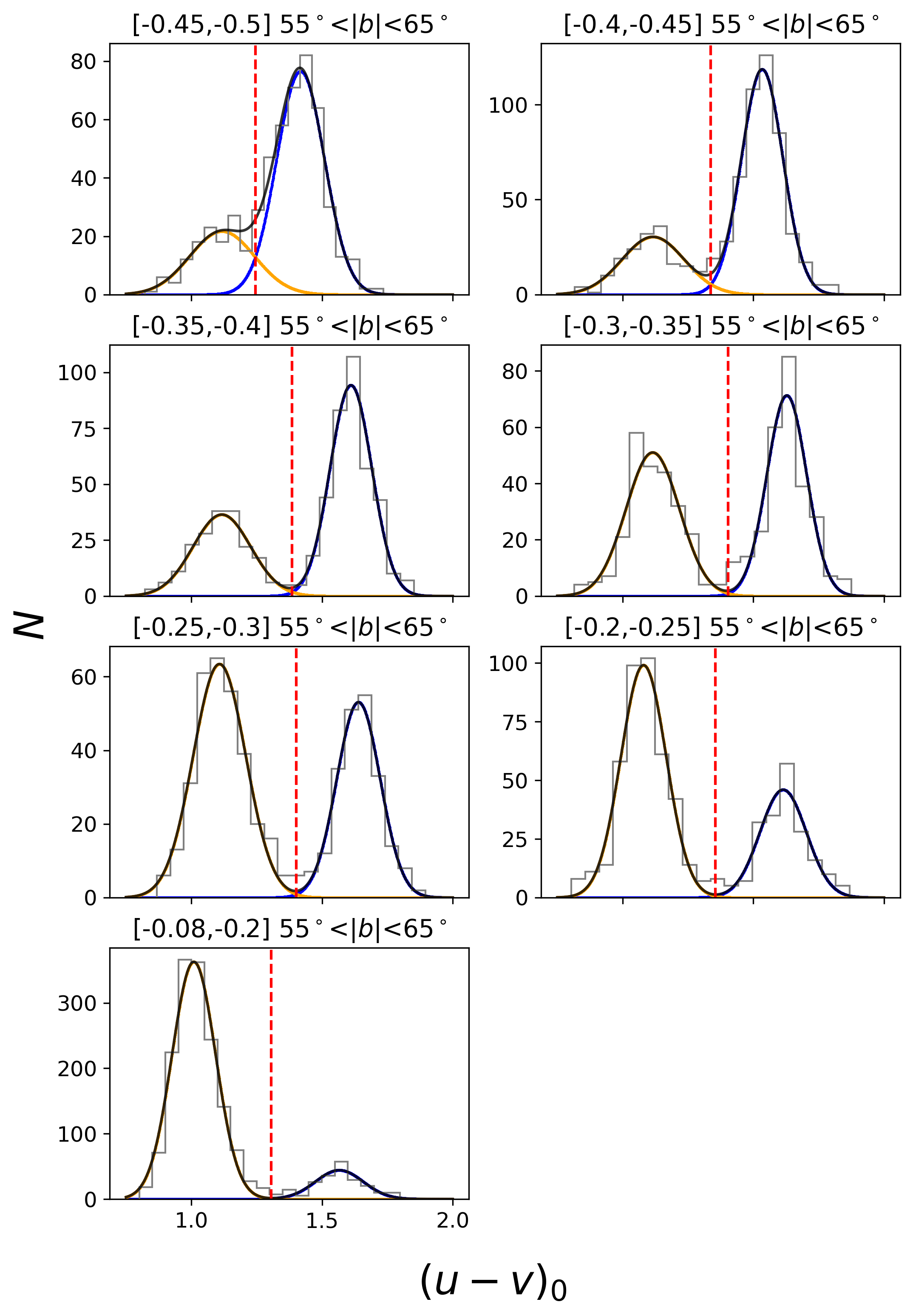}
    \caption{Similar to Figure\,~\ref{Fig:3}, but for the stars with 55$^\circ$ $<$ $\lvert b \rvert $ $<$ 65$^\circ$.}
    \label{fig:side:a1}
  \end{minipage}%
   \hspace{0.3in}
  \begin{minipage}[t]{0.5\linewidth}
    \centering
    \includegraphics[scale=0.5]{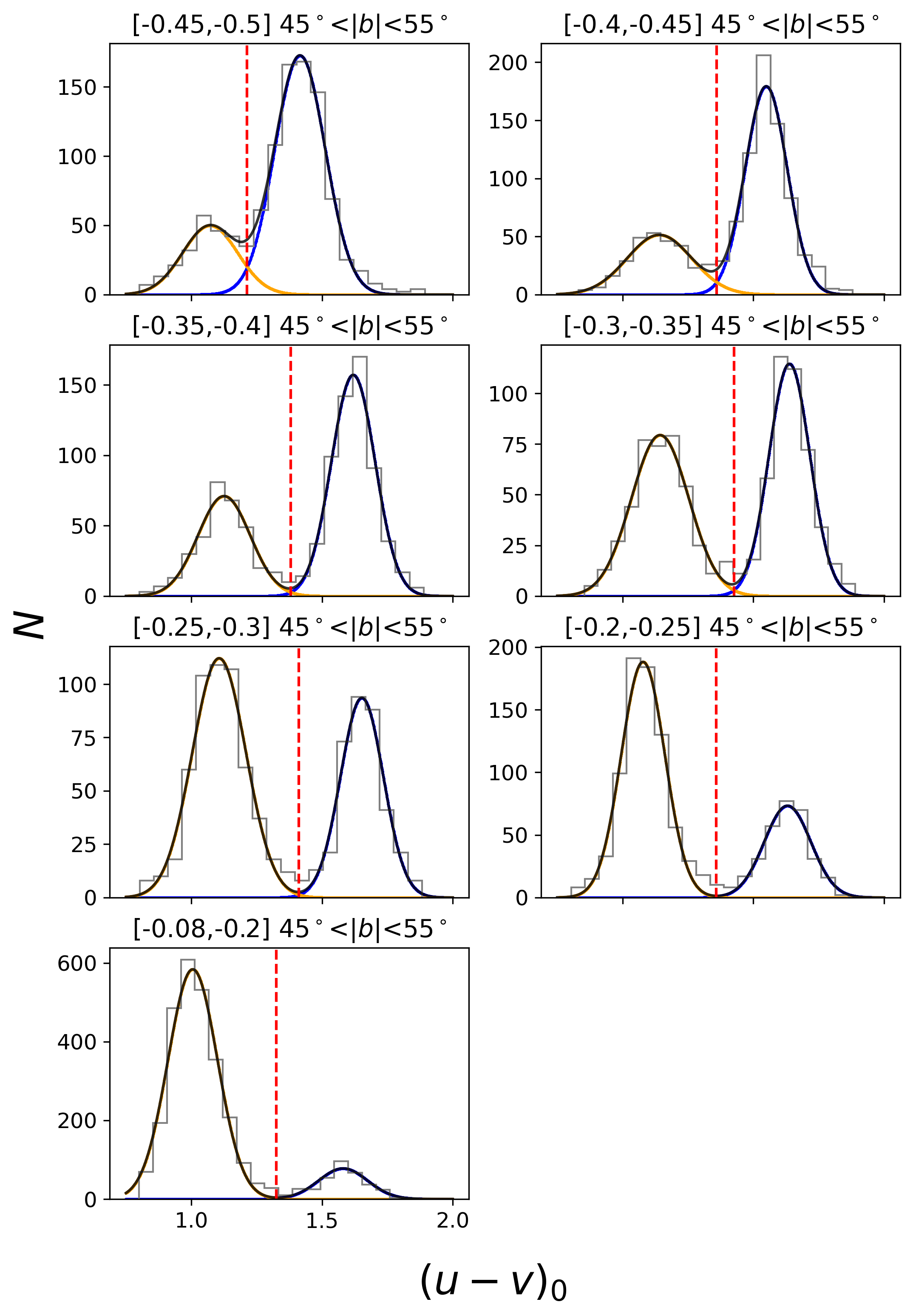}
    \caption{Similar to Figure\,~\ref{Fig:3}, but for the stars with 45$^\circ$ $<$ $\lvert b \rvert $ $<$ 55$^\circ$.}
    \label{fig:side:b1}
  \end{minipage}
\end{figure}

\begin{figure}
  \begin{minipage}[t]{0.5\linewidth}
    \centering
    \includegraphics[scale=0.5]{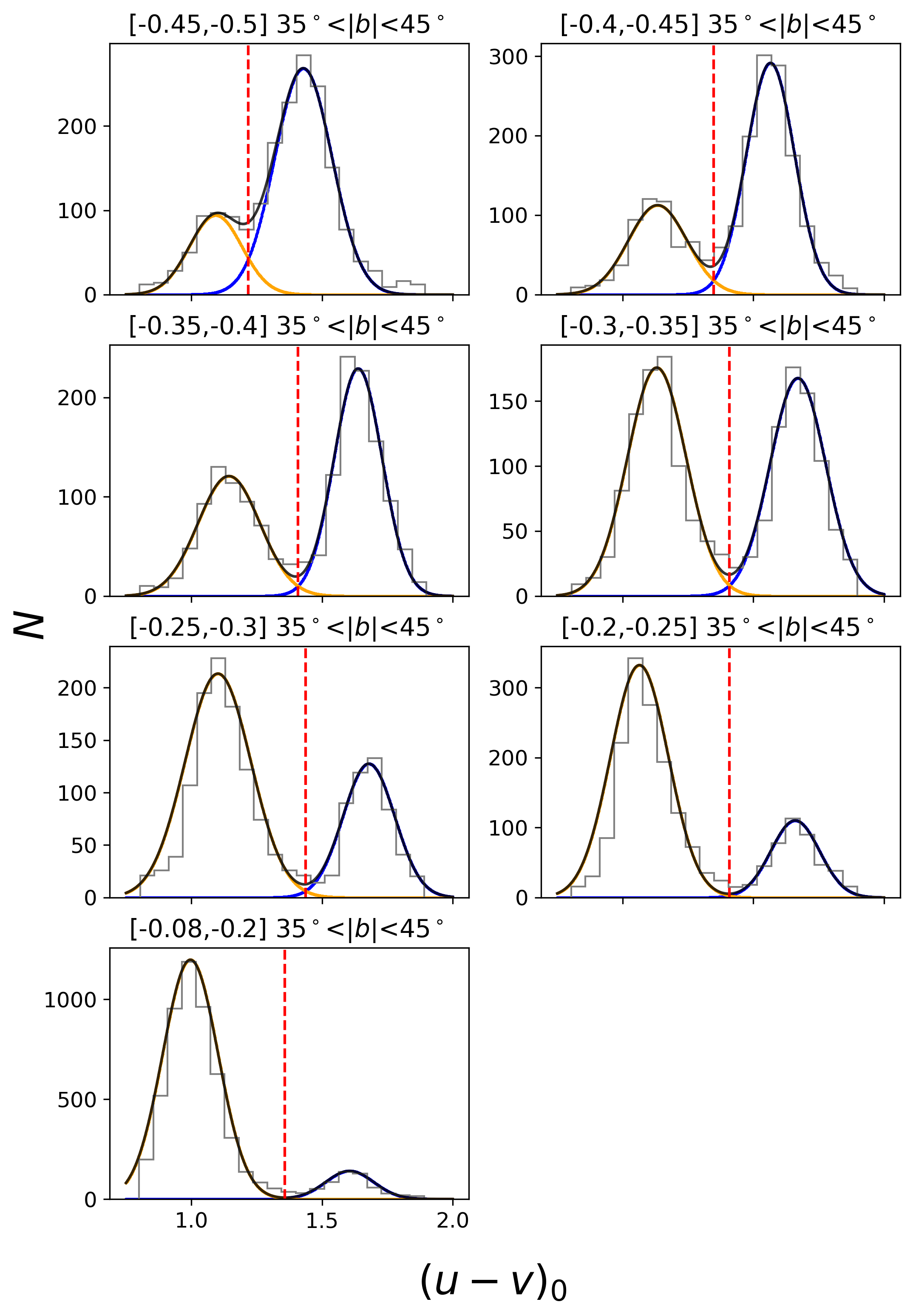}
    \caption{Similar to Figure\,~\ref{Fig:3}, but for the stars with 35$^\circ$ $<$ $\lvert b \rvert $ $<$ 45$^\circ$.}
    \label{fig:side:a2}
  \end{minipage}%
   \hspace{0.3in}
  \begin{minipage}[t]{0.5\linewidth}
    \centering
    \includegraphics[scale=0.5]{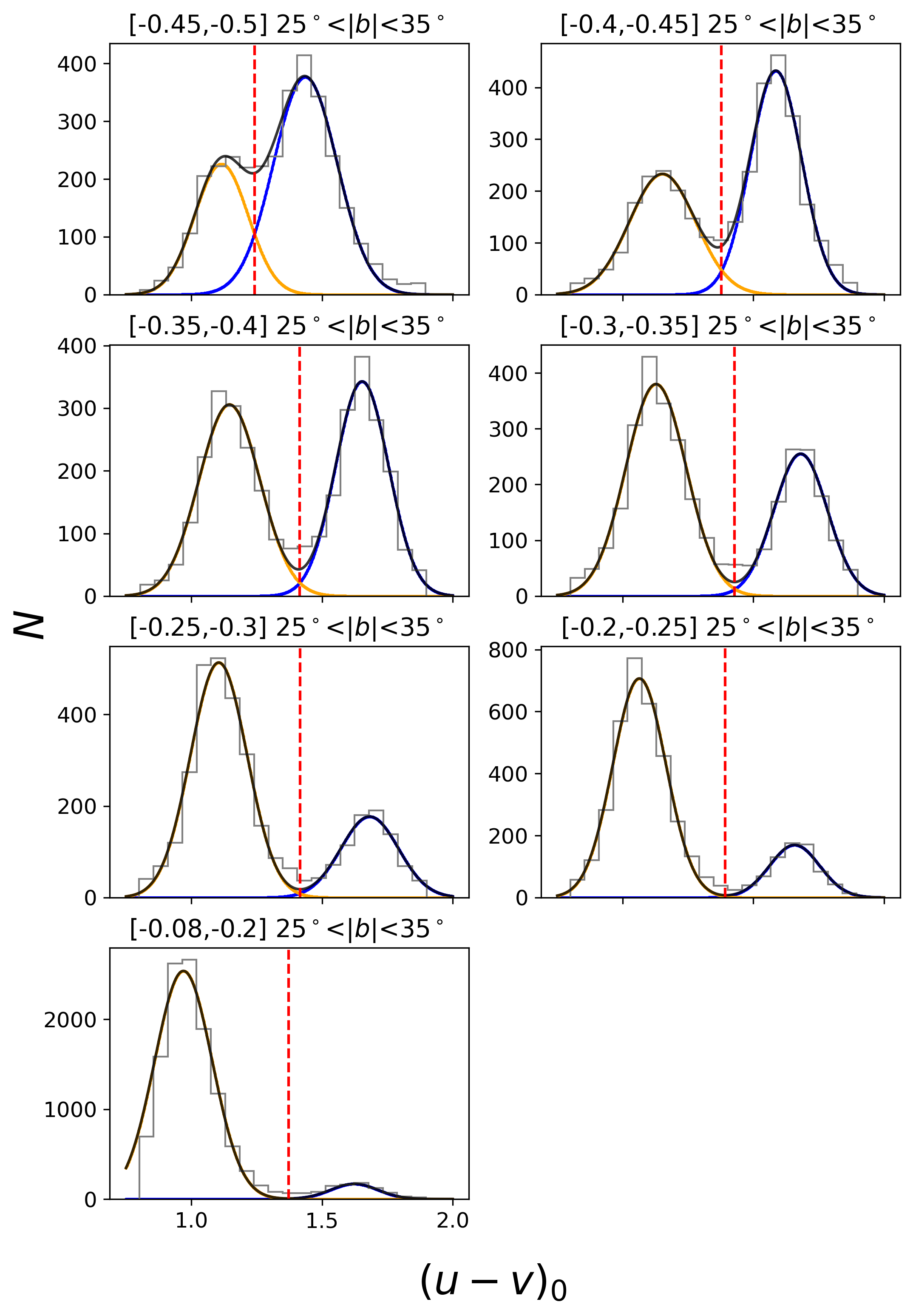}
    \caption{Similar to Figure\,~\ref{Fig:3}, but for the stars with 25$^\circ$ $<$ $\lvert b \rvert $ $<$ 35$^\circ$.}
    \label{fig:side:b2}
  \end{minipage}
\end{figure}

\begin{figure}
  \begin{minipage}[t]{0.5\linewidth}
    \centering
    \includegraphics[scale=0.5]{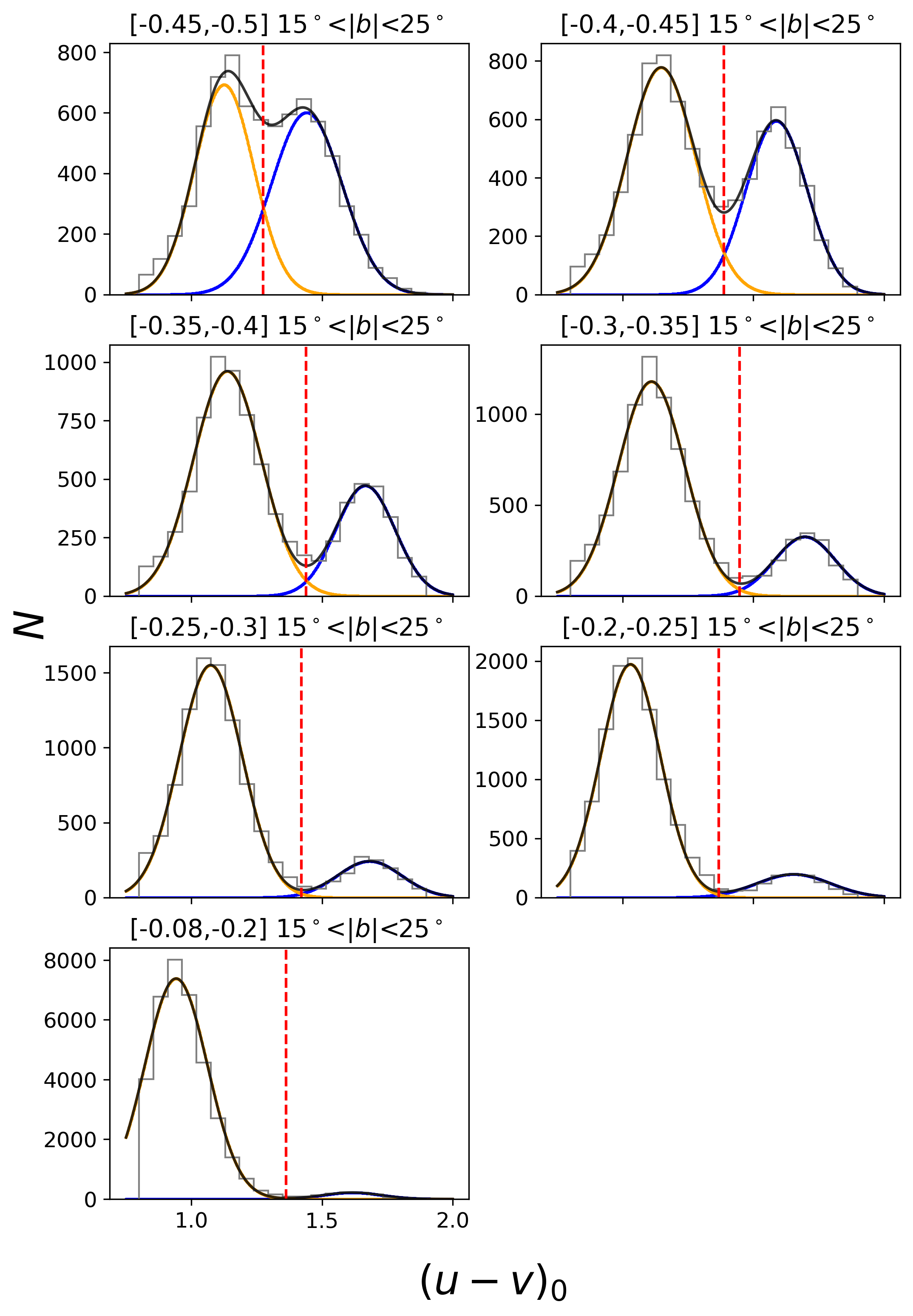}
    \caption{Similar to Figure\,~\ref{Fig:3}, but for the stars with 15$^\circ$ $<$ $\lvert b \rvert $ $<$ 25$^\circ$.}
    \label{fig:side:a3}
  \end{minipage}%
   \hspace{0.3in}
  \begin{minipage}[t]{0.5\linewidth}
    \centering
    \includegraphics[scale=0.5]{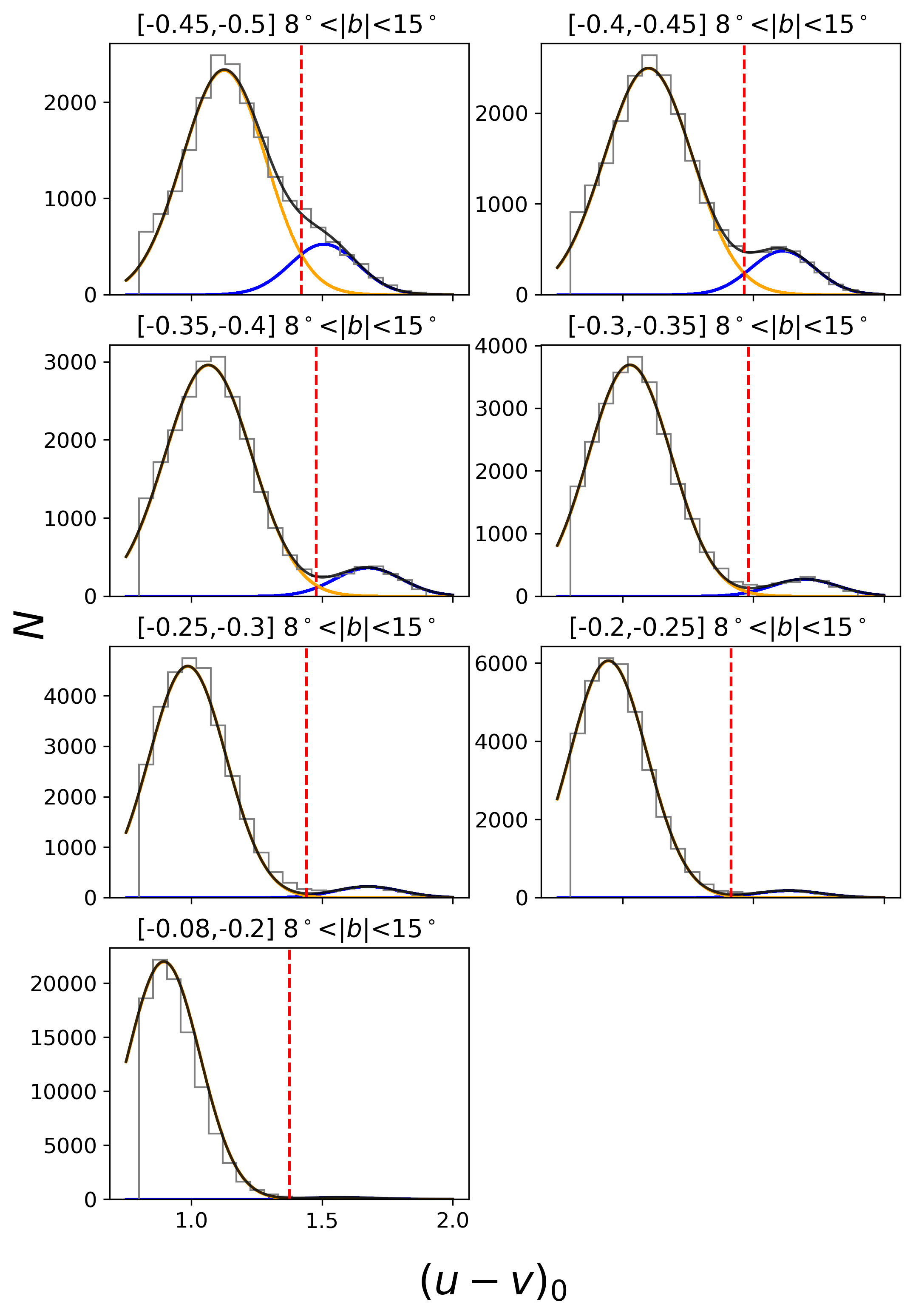}
    \caption{Similar to Figure\,~\ref{Fig:3}, but for the stars with 8$^\circ$ $<$ $\lvert b \rvert $ $<$ 15$^\circ$.}
    \label{fig:side:b3}
  \end{minipage}
\end{figure}

\clearpage

\end{CJK*}	
\end{document}